\documentclass[preprint2]{aastex}
\usepackage[]{natbib}
\usepackage{spr-astr-addons}
\usepackage{cellspace}
\def\vk{v_{\rm K}}           
\def\betiabs{|\beta_{\rm i0}|}
\def\betio{\beta_{\rm i0}}

\def\upo{\Upsilon_{\rm 0}}
\def\vk{v_{\rm K}}           

\newcounter{mypapers}

\shorttitle{Star and planet formation: magnetic fields}
\shortauthors{Salmeron R.}

\begin{document}


\title{Formation of stars and planets:\\
 the role of magnetic fields}


\author{R. Salmeron} 
\affil{Research School of Astronomy \& Astrophysics, Research School of Earth Sciences and
Planetary Science Institute, 
The Australian National University, 
Cotter Road
Weston Creek, ACT 2611
AUSTRALIA
}

%
%
%
%


\begin{abstract}
Star formation is thought to be triggered by gravitational collapse of the dense cores of molecular clouds. Angular momentum conservation during the collapse results in the progressive increase of the centrifugal force, which eventually halts the inflow of material and leads to the development of a central mass surrounded by a disc. In the presence of an angular momentum transport mechanism, mass accretion onto the central object proceeds through this disc, and it is believed that this is how stars typically gain most of their mass. However, the mechanisms responsible for this transport of angular momentum are not well understood. Although the gravitational field of a companion star or even gravitational instabilities (particularly in massive discs) may play a role, the most general mechanisms are \emph{turbulence viscosity} driven by the magnetorotational instability (MRI), and \emph{outflows} accelerated centrifugally from the surfaces of the disc. Both processes are powered by the action of magnetic fields and are, in turn, likely to strongly affect the structure, dynamics, evolutionary path and planet-forming capabilities of their host discs. The weak ionisation of protostellar discs, however, may prevent the magnetic field from effectively coupling to the gas and shear and driving these processes. Here I examine the viability and properties of these magnetically-driven processes in protostellar discs. The results indicate that, despite the weak ionisation, the magnetic field is able to couple to the gas and shear for fluid conditions thought to be satisfied over a wide range of radii in these discs.  
\end{abstract}





\section{Introduction}

The birth of stars, a process so commonly occurring throughout the whole universe, is as puzzling as it is fascinating.
The current paradigm for star formation, framed originally by Kant and 
Laplace in the $18^{\rm th}$ century, suggests that stars are born via gravitational collapse of the dense cores of molecular clouds, in turn condensed out of the diffuse 
interstellar medium.  Unfortunately, this compelling visualisation and its physical 
underpinning 
are in apparent contradiction.

The reason for the above statement is that conservation of angular momentum during the collapse results in the progressive increase of the centrifugal force, which eventually halts the 
infalling gas and leads to the development of a central mass (i.e.~a `protostar') surrounded by a flattened disc of material (an `accretion disc').    Observational evidence for the presence of such discs around young stellar objects is compelling. It comprises imaging in the near infrared and optical wavebands \citep[e.g.~see the reviews by][]{WSWM07, MSC00} as well as interferometric studies that have resolved the velocity profile and structure in the inner regions of discs, up to $\sim$ a few tens of AU from the centre \citep[e.g.][]{DGH07, WL00}. This phenomenon is, in fact, not limited to young stars. Many astrophysical systems exhibit the characteristic disc-like structure that naturally results when inward motion in the plane of rotation is restricted by angular momentum conservation, while collapse continues in the perpendicular (polar) direction. Such structures are commonly associated, for example, with the discs of material feeding the cores of active galaxies and black holes. 

The difficulty in progressing past this stage is clear when we recall that these protostellar discs are dynamically stable.  A typical disc is differentially rotating, with an angular momentum ($L$) profile that increases with radius ($dL/dr > 0$). 
 Like our own planet, its material is unable to move closer to the central object unless it loses angular momentum. This can only be accomplished by redistributing it to other parts of the disc, or by transferring it away from the star-disc system altogether. Indeed, the efficiency and evolution of this phase of the star formation process is regulated by the rate at which angular momentum can be transferred away from the accreting matter \citep[e.g.][]{AL93}. Fortunately for us, the Earth is unable to do so (and stays in its orbit!) but, how does nature overcome this orbital stability, if stars are to form in the first place?

The mechanism(s) responsible for generating and sustaining this transport of angular momentum are not well understood. Different options have been invoked by many authors over several decades, with various degrees of success. It is known, however, that the molecular viscosity of accretion discs is far too low to explain the inferred accretion rates of these objects (\citealt{P81}; see also \citealt{FKR92}). Specifically, viscous diffusion is able to propagate disturbances in timescales $\sim l^2/\nu$, where $l$ is the distance the disturbance travels as a result of the action of the kinematic viscosity $\nu$. Taking $l \sim 10^{10}$ cm and $\nu = 10^5$ cm$^2$ s$^{-1}$, this timescale is $\sim 3 \times 10^7$ years \citep{balbus03b}, far too long to explain the variability observed in some accreting systems. Gravitational instabilities \citep[e.g.][]{ARS89, D07a} may be a viable transport process, but this mechanism may be effective mainly in the outer regions of massive discs 
or at early stages of the star formation process, when the mass of the disc is a relatively large fraction of that of the central object. 
Other proposed mechanisms are not general enough (i.e.~the presence of a companion star) or do not work at all (i.e.~convection). In recent times it has been realised that the answer may lie in the complex interaction of gas and magnetic fields present in the disc \citep[][]{BH91, BH98}. Magnetic activity, in other words, may be the critical ingredient in the recipe nature follows to assemble new stars. 

From the above discussion, it is clear that angular momentum lies at the core of accretion-disc dynamics, and to understand how it is transported it is crucial to pay close attention to the 
physical conditions of the fluid. Most analytical and numerical studies conducted so far have adopted a number of simplifications to treat the fluid, and the interactions of matter with the magnetic field, in accreting systems. 
 In the cold, dense environments typical of protostellar discs,\footnote{Note that the terms `protostellar' and `protoplanetary' discs are used interchangeably in the text.} it is particularly important to adequately treat the slippage between the magnetic field and the gas (the \emph{magnetic diffusivity}) that results from the weak ionisation of the disc matter. 

This paper is organised as follows. Section \ref{sec:B} briefly discusses the key ways in which magnetic fields may impact on disc physics, paying particular attention to the concept and types of field-matter diffusivity. Section \ref{sec:ang} discusses the `angular momentum transport problem' in accretion discs and introduces the most promising mechanisms to transfer it away, enabling accretion to proceed: the action of the magnetorotational instability \citep[hereafter MRI;][]{BH91, BH98} and the acceleration of outflows from the disc surfaces \citep{KP00, POFB07, KS11}. These mechanisms are described in more detail in Sections \ref{sec:disk-wind} and \ref{sec:MRI}, respectively; and Section \ref{sec:combined} discusses the possibility of their joint operation. Section \ref{sec:planets} briefly summarises the main points discussed in the paper and concludes. 

\section{Magnetic activity in protostellar discs}
\label{sec:B}
 
Protostellar discs are known to be magnetised. Evidence of this comes from different sources. On the one hand, there is strong evidence for an enhanced magnetic activity in young stellar objects \citep[e.g.~see the reviews by][]{FTGS07, GFM00}. As these authors point out, it is likely that the strong magnetic field near the stellar surface, extends out to the circumstellar disc. On the other hand, the remnant magnetisation of primitive meteorites suggests that magnetic fields were important in our own solar nebula as well \citep{LS78}. 

This property of discs is important for the present topic because
magnetic fields are thought to play key roles in the dynamics and evolution of protostellar discs. As mentioned above, they redistribute angular momentum via magnetohydrodynamic (MHD) turbulence \citep{SGBH00}, as well as by magnetocentrifugal winds accelerated from the disc surfaces and also possibly from the star/disc magnetosphere \citep[e.g.~see the reviews by][]{KS11, KP00, POFB07}. Additionally, magnetically-driven mixing strongly influences the chemistry of the disc \citep[e.g.][]{SWH06, IN06} and critically affects the dynamics, aggregation and overall evolution of dust particles \citep{TWBY06, Ci07} -- the assembly blocks of planetesimals according to the `core accretion' model of planet formation \citep{P96}. Furthermore, magnetic fields can also modify the response of the disc to gravitational perturbations introduced by forming planets, in turn influencing the rate and even the direction of planetary migration through the disc \citep{T03, FTN05a, MMI08, JGM06}. 
Finally, jets accelerated via magnetic stresses may be the sites of \emph{chondrule} formation \citep[primitive, thermally-processed pieces of rock found in meteorite samples; see e.g.][]{Uesugi05a, DMCB10}; and magnetically-driven activity near the disc surface can produce a hot, tenuous corona, \citep[e.g.][]{FS03} and influence the observational signatures of these objects. 

In protoplanetary discs, however, the magnetic diffusivity can be severe enough to limit -- or even completely inhibit -- the above-mentioned processes. The particular role magnetic fields are able to play in these environments is, therefore, largely determined by the degree of coupling between the field and the neutral gas. This important topic is discussed next.

\subsection{Magnetic diffusivity}
\label{subsec:Magdiff}

 \begin{figure*}[tbp]
\centering
\includegraphics[width=3.72in]{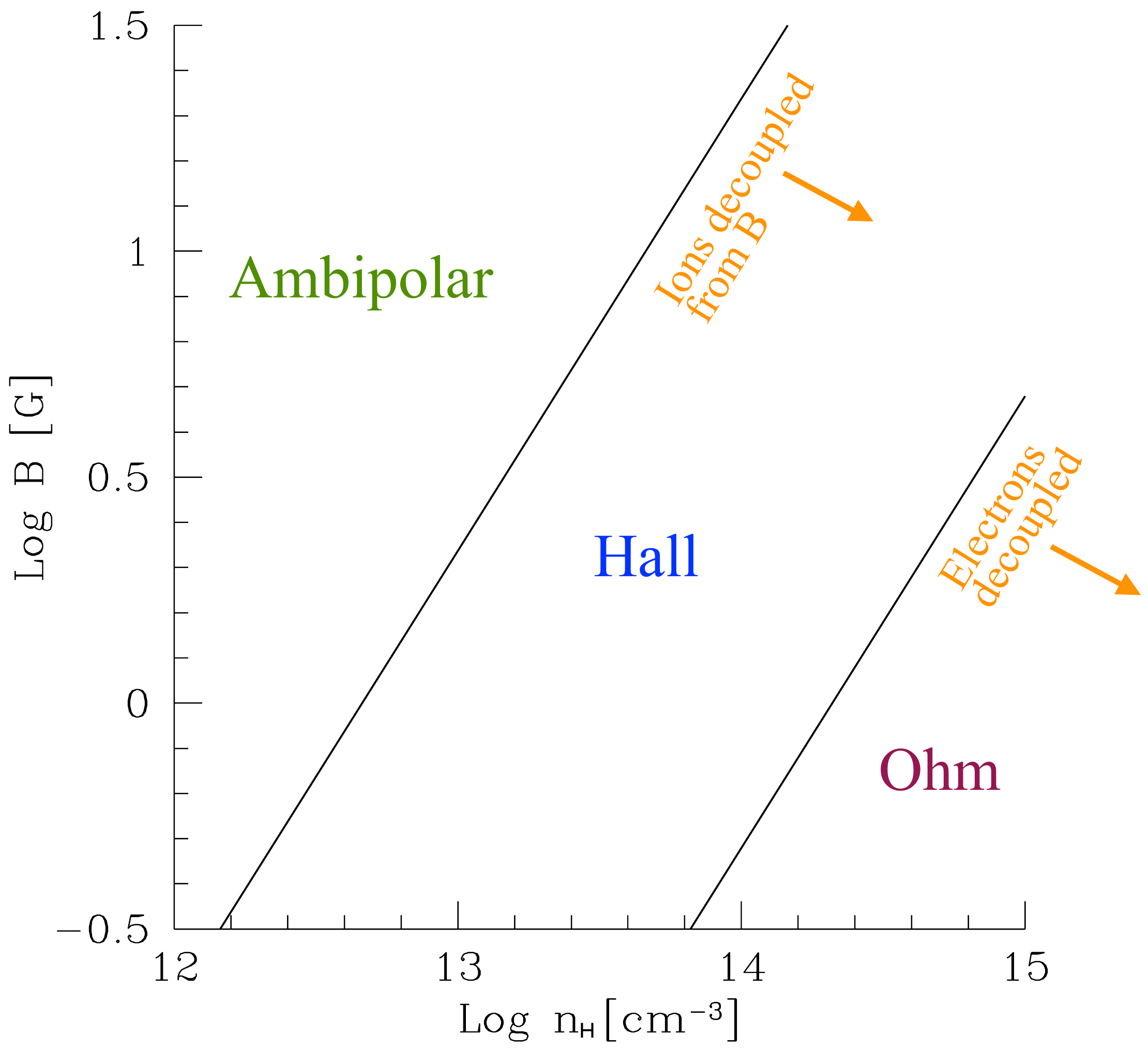}
\caption{Magnetic diffusivity regimes -- Ambipolar, Hall and Ohm --  in a log $n_{\rm H}$ [cm$^{-3}$] -- log $B$ [G] plane for $T = 280$ K when the charged particles are ions and electrons only, in weakly-ionised gas \citep[see also][]{W07}. The different regions are delineated by the values of the ion and electron Hall parameters $|\beta_{\rm i}|$ and $|\beta_{\rm e}|$, respectively. The $\beta_j$ measure the ratio of the gyrofrequency to the collision frequency of charged species $j$ with the neutrals and are, therefore, a measure of the relative importance of the Lorentz and drag forces on the motion of the charged species (see text for further details). The ions (electrons) decouple from the magnetic field when $|\beta_{\rm i}|$ ($|\beta_{\rm e}|$) drops below unity. This occurs to the right of the appropriate inclined line in the figure. 
}
\label{fig:diff}
\end{figure*}

The electrical conductivity of a fluid is determined by the abundance and drift motion of the charged species embedded in it. These include, in general, ions, electrons and charged dust grains. The abundance of these charged species -- measured by the ionisation fraction -- results, in turn, from the balance of ionisation and recombination processes acting in the system under consideration. In protostellar discs, in particular, thermal ionisation is ineffective outside the central $\sim 0.1$ AU around the protostar, as the temperature drops below $\sim 1000$ K. In the bulk of the disc, therefore, the most active ionising agents are stellar X-ray and UV radiation, interstellar cosmic rays (see e.g.~\citealt{H81, SWH04, GKS05}), and -- to a much lesser extent -- the decay of radioactive elements present within the disc \citep[primarily $^{40}$K; see][]{UN81, UN90}. The attenuation column of X-rays and cosmic rays is 10  and 100 g cm$^{-2}$, respectively. In contrast, according to the minimum-mass solar nebula model \citep{H81}, the surface density of the disc at 1 AU is $\sim$ 1700 g cm$^{-2}$. Therefore, the resulting fractional ionisation in the disc interior, specially for radii $\lesssim 10$ AU, is unlikely to be high enough to provide good coupling between the charged and neutral components of the fluid over the entire cross-section of the disc. 

The degree of coupling between the charged species and the neutral gas in diffusive media is typically measured by the Hall parameter $\beta_j$, the ratio of the gyrofrequency to the collision frequency of species $j$ with the neutrals. Here, we specialise to ions and electrons, and denote them by subscripts $i$ and $e$, respectively\footnote{\label{foot:note2} For charged species $j$, endowed with mass $m_j$ and charge $|Z_j| e$, $\beta_j =  (|Z_j| e B/m_j c) (1/\gamma_j \rho)$. The first parenthesis is the gyrofrequency around a magnetic field of {\em signed} magnitude $B = |\textbf{B}| \ {\rm sgn}\{B_z\}$ (so that the $\beta_j$ retain their dependence  on the magnetic field polarity); and $c$ is the speed of light. The second parenthesis measures the collision frequency of species $j$ with the neutral gas, of density $\rho$. The quantity $\gamma_j = \langle \sigma v\rangle_j/(m_j + m)$, where $\langle \sigma v\rangle_j$ is the rate coefficient of momentum exchange of species $j$ with the neutrals, and $m = 2.33 \ m_{\rm H}$ is the mean mass of the neutral particles in units of the hydrogen mass.}. This parameter measures the relative importance of the  Lorentz and drag forces on the motion of the charged species. The magnetic diffusivity is a tensor \citep{C76, NH85, NU86a, W99} whenever the gyrofrequency of some charge carriers is larger than their frequency of collisional exchange of momentum with the neutrals (or $|\beta| > 1$ for at least some of the charged species). This is expected, and just reflects the anisotropy that the magnetic field is able to impart on the drift of the charges when at least some of these are well coupled to it. In such environments, the diffusivity tensor components are the Pedersen ($\eta_{\rm P}$), Hall ($\eta_{\rm H}$) and Ohm ($\eta_{\rm O}$) terms. We also often use the ambipolar diffusivity term, given by $\eta_{\rm A} = \eta_{\rm P} - \eta_{\rm O}$. When the charge carriers are ions and electrons, it can be shown that the following relations apply \citep{W07}: $\eta_{\rm H} = \beta_{\rm e} \eta_{\rm O}$ and $\eta_{\rm A} = \beta_{\rm i} \beta_{\rm e} \eta_{\rm O}$. Note also that $\eta_{\rm O}$, $\eta_{\rm A}$ and $\eta_{\rm P}$ are always positive (as $\eta_{\rm O}$ is independent of $B$, whereas $\eta_{\rm A}$ depends on $B^2$), but $\eta_{\rm H}$ scales linearly with $B$ and can, therefore, be either positive or negative depending on the field polarity (e.g.~the sign of $B_z$). 

The relative drifts of the charged species with respect to the bulk of the neutral gas delineate three different conductivity regimes, as detailed below.

\begin{itemize}
\item \emph{Ambipolar diffusion limit}. In this regime, which  is important at relatively low densities, most charged species are primarily tied to the magnetic field through electromagnetic stresses (e.g.~$|\beta_{\rm e}| \gg |\beta_{\rm i}| \gg 1$, or $\eta_{\rm A} \gg |\eta_{\rm H}| \gg \eta_{\rm O}$).  This implies that the magnetic field is effectively frozen into the ionised component of the fluid and drifts with it through the neutrals. This diffusion mechanism is thought to be an important contributor to the loss of magnetic support in molecular clouds, facilitating the onset of cloud collapse. In protostellar discs, this regime is expected to be dominant at radial distances beyond $\sim 10$ AU and near the surface closer in.

\item \emph{Resistive (Ohmic) limit}, in which the ionised species are mainly tied to the neutrals via collisions ($1 \gg |\beta_{\rm e}| \gg |\beta_{\rm i}|$, or $\eta_{\rm O} \gg |\eta_{\rm H}| \gg \eta_{\rm A}$). This regime is predominant in high-density environments, typically close to the midplane in the innermost regions of protostellar discs ($R \sim 0.1$ -- $1$ AU), just outside the region in the vicinity of the protostar where thermal ionisation is effective. In this limit, the collision frequency of all the charged species with the neutrals is high enough to effectively suppress the drift of the former through the bulk of the gas.  

\item \emph{Hall limit}, characterised by a varying degree of magnetic coupling amongst charged species. This  regime is important at intermediate densities, in-between those at which the ambipolar diffusion and resistive limits are dominant. In this regime, some charged species (typically electrons) are still tied to the magnetic field whereas more massive particles (such as ions and charged dust grains) have already decoupled to it, and are collisionally tied to the neutral gas ($|\beta_{\rm e}| \gg 1 \gg |\beta_{\rm i}|$). In this limit, therefore, $|\eta_{\rm H}| \gg \eta_{\rm A}$ and $\eta_{\rm O}$. As discussed above, when this regime dominates, the magnetic response of the disc is no longer invariant under a global reversal of the magnetic field polarity (e.g.~\citealt{WN99, KSW10}, hereafter KSW10). The Hall regime is expected to dominate over fluid conditions satisfied over an extended range of radii in protostellar discs \citep[e.g.][]{SS02a, SS02b} 
\end{itemize}

In the collisionally-dominated resistive limit, the resulting magnetic diffusivity is a scalar, the familiar Ohmic resistivity. This is simply because collisions occur in all directions, therefore the impulse they communicate to the charges is randomly oriented. In the other limits, however, where at least some charged species are well tied to the magnetic field, the diffusivity is a tensor.

Previous studies on the magnetic activity of discs (e.g.~\citealt{W99, SS02a, SS02b, SW03, SW05, WS11}, hereafter WS11) have highlighted the importance of incorporating in these studies all the three field-matter diffusion mechanisms described above, as their relative importance is a strong function of location within the disc. This is illustrated in Fig.~\ref{fig:diff}, which shows the regions where the ambipolar, Hall and Ohmic diffusivity terms dominate in a Log $B$ -- Log $n_{\rm H}$ plane \citep[e.g.][]{W07}, assuming that the charged particles are ions and electrons, in a weakly-ionised plasma. As expected, ambipolar diffusion is dominant at low densities and strong fields whereas the opposite is true for the Ohmic regime. The large, intermediate region of parameter space between these two limits is dominated by the Hall diffusivity.  For example, for $r = 1$ AU the number density at the disc midplane, according to the minimum--mass solar nebula model \citep{H81}, is $\sim 6 \times 10^{14}$ cm$^{-3}$. This implies that, under the adopted assumptions, the gas at this location would be in the Ohmic diffusivity regime for field strengths $ \lesssim 2.5$ G and in the Hall limit for stronger fields. 

More generally, the location of the curves delineating  the regions where each of the diffusivity regimes dominates is a function of the mass of the charged particles and the fractional ionisation of the fluid. If the charged particles are more massive, the Ambipolar-Hall and Hall-Ohm transitions will occur at stronger fields for a given value of the neutral density. Similarly, if the ionisation fraction increases (e.g. the neutral density is a lower ratio of the total density of the fluid), the above transitions will take place at comparatively lower magnetic field strengths. A particularly important scenario is when dust particles are well mixed with the gas.

The degree of field--matter coupling is strongly dependent on the abundance and size distribution of dust particles mixed with the gas. Dust grains tend to dramatically reduce the fractional ionisation of the fluid, as charged particles can stick to them and recombine on their surfaces. Additionally, grains can also become an important species in high-density regions \citep[e.g.][]{UN90, NNU91}. As they have relatively  large cross sections, grains typically become decoupled from the magnetic field at lower densities than those for which other --smaller-- species do. As a result, the conductivity of the fluid diminishes when dust particles (particularly when these are small) are well mixed with the gas. This is most relevant at early stages of the accretion process, or when turbulent motions prevent the grains from settling towards the disc midplane. In fact, recent diffusivity calculations for a minimum-mass solar nebula disc at 1 AU (\citealt{W07}; WS11) have shown that  the magnetic coupling may be adequate over the entire disc cross-section, provided that dust grains are settled out of the gas phase. 

A word of caution applies to the above statements. The turbulent stirring of dust grains is, evidently, dependent on the disc being able to sustain turbulence when dust grains are well mixed with the gas. However, as disc turbulence is thought to be a magnetically-driven process (see Section \ref{sec:MRI}), the reduction in the degree of field-matter coupling mediated by dust grains may, in turn,  weaken the turbulent motions, or even prevent their development in dusty plasmas. Disc dynamics is then the result of a complex interplay: When dust grains are well mixed with the gas, the disc may not be susceptible to turbulence, and the grains may efficiently settle to the midplane. On the other hand, once the grains are effectively removed from the gas, turbulence may be viable again, and the turbulent motions may stir the grains back into the gas phase. The resulting structure and evolution of discs, incorporating this important feedback effect, have not yet been calculated self-consistently. 
 
\section{Angular momentum transport}
\label{sec:ang}

Most current models of accretion discs are viscous models that adopt, in some form, the $\alpha$-prescription of \citet{SS73}. In this formulation, the radial-azimuthal component of the stress tensor is assumed to scale with the gas pressure, or $w_{r \phi} = \alpha P$. The associated turbulent viscosity is parameterised as $\nu_{t} = \alpha c_{\rm s} h_{\rm T}$, where $c_{\rm s}$ is the isothermal sound speed and $h_{\rm T}$ is the tidal density scale-height of the disc (e.g.~resulting from the vertical gravitational compression). Despite its usefulness in providing an operational framework  to support the modelling of accretion discs, this methodology does not offer any explanation for the physical nature of the accretion torque, whose origin remains unspecified. In fact, in this formulation, all the unknowns and uncertainties relating to the accretion stress are just effectively lumped into the convenient parameter $\alpha$.

In general, one of the most natural ways of generating turbulence is via hydrodynamical instabilities. Laboratory shear flows, for example, readily break into turbulence at sufficiently high Reynolds\footnote{The Reynolds number $R_{\rm e} \equiv vl/\nu$, where $v$ is the flow velocity, $\nu$ is the kinematic viscosity and $l$ is an adopted length scale of the flow. $R_{\rm e}$ then represents the ratio of the inertial to viscous forces in the fluid.} numbers. However, Keplerian discs\footnote{In a Keplerian disc the orbital velocity is given by the relation $v_{\rm K} = \sqrt{GM/r}$, where $G$ is the gravitational constant, $M$ is the mass of the central object and $r$ is the cylindrical radius. The ratio $\Omega_{\rm K} = v_{\rm K}/r$ is the Keplerian angular frequency.} satisfy the Rayleigh's hydrodynamical stability criterion (angular momentum increases with radius), so the generation and sustaining of hydrodynamic turbulence in these systems is, at best, unproven \citep[e.g.~see review by][]{BH98}. Convective turbulence has also been considered as an option \citep{LP80, LPK93}, but further studies appear to indicate that this mechanism transports angular momentum towards the central object \citep{CP92, RG92, C96, SB96}. More generally, the sign of the radial flux generated by convective turbulence may depend on the ratio of the epicyclic frequency and the wave frequency of the fluid \citep{balbus03b}: Angular momentum is likely to be transported inwards by long-period, incompressible perturbations; and outwards by high-frequency, compressible disturbances. Still, other authors have pointed out that hydrodynamic waves originated by the gravitational field of a companion star could possibly transport angular momentum \citep{VD89, RS93}. This mechanism, however, makes use of an ÔexternalÕ source to excite, and maintain, such density waves; therefore it can not explain accretion in stars that do not belong to multiple systems. Finally, although gravitational instabilities \citep[e.g.][]{ARS89, D07a} may play a role, these may be effective mostly during early stages of evolution of these objects, and for particularly massive discs. 

It is increasingly clear that the two most generally applicable angular momentum transport processes are mediated by the action of magnetic fields and their interaction with disc matter. They are the following.

\begin{enumerate}

\item \emph{Vertical transport via winds} accelerated centrifugally from the disc surfaces (`disc winds'; e.g. see the reviews by \citealt{KP00, POFB07, KS11}). This notion is strengthened by the ubiquity of these outflows in star forming regions.

\item \emph{Radial transport through turbulent viscosity} induced by the magnetorotational instability (MRI; \citealt{BH91, BH98}). The MRI, essentially, taps into the free energy contributed by the differential rotation of the disc, and converts it into turbulent motions that transfer angular momentum radially outwards.

\end{enumerate}

These processes are, in turn, likely to strongly affect the structure, dynamics, evolutionary path and planet-forming capabilities of their host discs. In the next two sections I describe in more detail the operation of these two forms of angular momentum transport in protostellar discs. I  then briefly consider the possibility for their operation at the same radius, but at different heights above the disc midplane.

 \section{Vertical transport via magnetocentrifugal outflows}
 \label{sec:disk-wind}
 
\begin{figure*}
\centering
\includegraphics[width=6.2in]{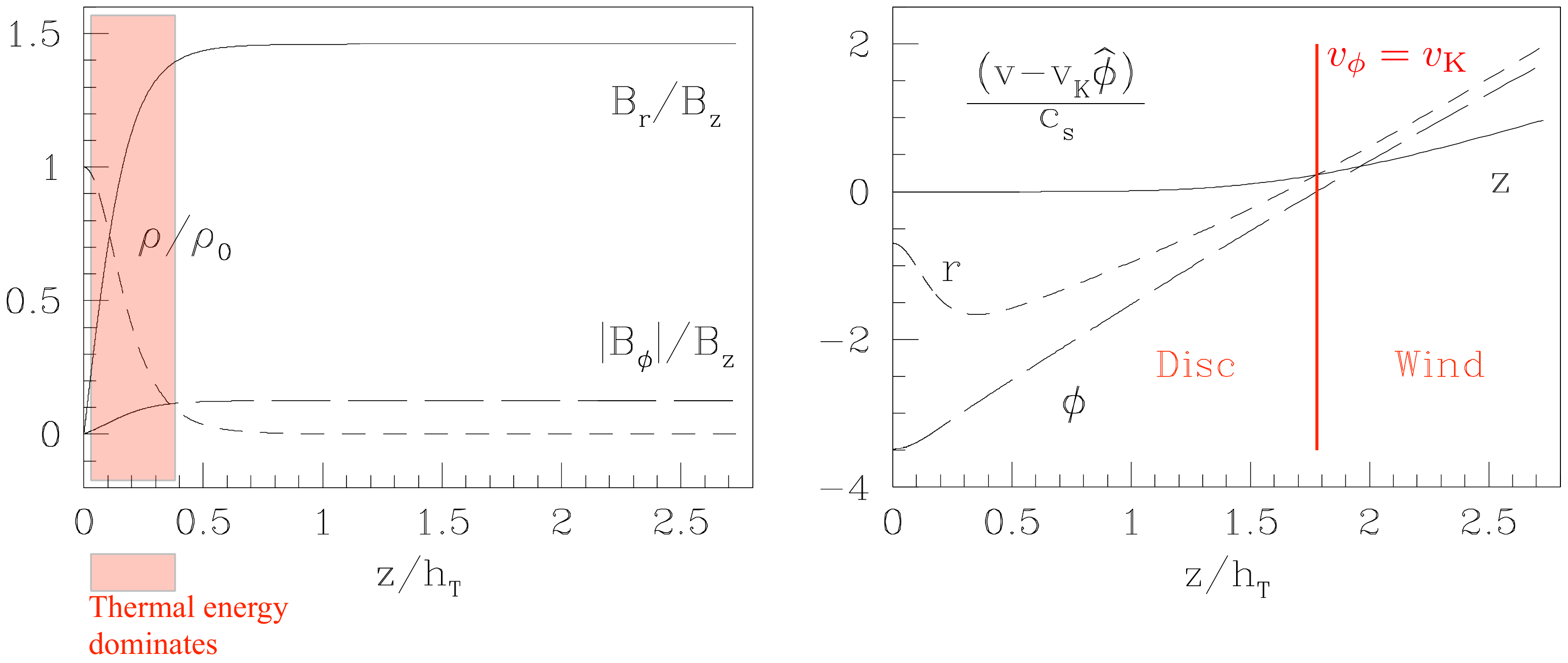}
\caption{Vertical structure of a strongly-magnetised, wind-driving disc solution, as a function of height above the disc midplane (in units of $h_{\rm T}$, the tidal density scale-height), under the assumption of pure ambipolar diffusivity. The model parameters are: $\Lambda = 10$, $a_{\rm 0} = 0.95$,  $\eta_{\rm P0} \Omega/c_{\rm s}^2= 0.09$, $\eta_{\rm H0}  \Omega/c_{\rm s}^2 = 0$, $\epsilon = 0.7$, $\epsilon_{\rm B} = 0$, and $v_{\rm K}/c_{\rm s} = 20$ (see text for details).  For simplicity, in this illustrative solution the diffusivity components are assumed to scale with height in such a way that the Elsasser number ($\Lambda$) remains constant with $z$. In a more general formulation, these parameters  would vary with height as dictated by the ionisation balance in the disc.   All curves terminate at the sonic point. 
\emph{Left}: Gas density ($\rho/\rho_{\rm 0}$, normalised by the midplane value) and magnetic field components (radial and azimuthal, $B_r/B_z$ and $|B_\phi|/B_z$, respectively; normalised by the vertical field $B_z$, which is constant with height). In the shaded region, the gas thermal energy dominates over the magnetic energy and the magnetic field lines are bent and sheared. \emph{Right}: Velocity components, with respect to the Keplerian velocity $v_{\rm K}$ and normalised by the isothermal sound speed $c_{\rm s}$. Within the disc (to the left of the red vertical line), the radial velocity is negative, the azimuthal velocity is sub-Keplerian and the vertical velocity is small. On the contrary, in the outflow region above this layer all three velocity components [$(\mathbf{v} - v_{\rm K}\hat{\mathbf{\phi}})/c_{\rm s}$] are positive and increasing with $z$.}
\label{fig:illus}
\end{figure*}

  \begin{figure*} 
\centering
\includegraphics[width=5.8in]{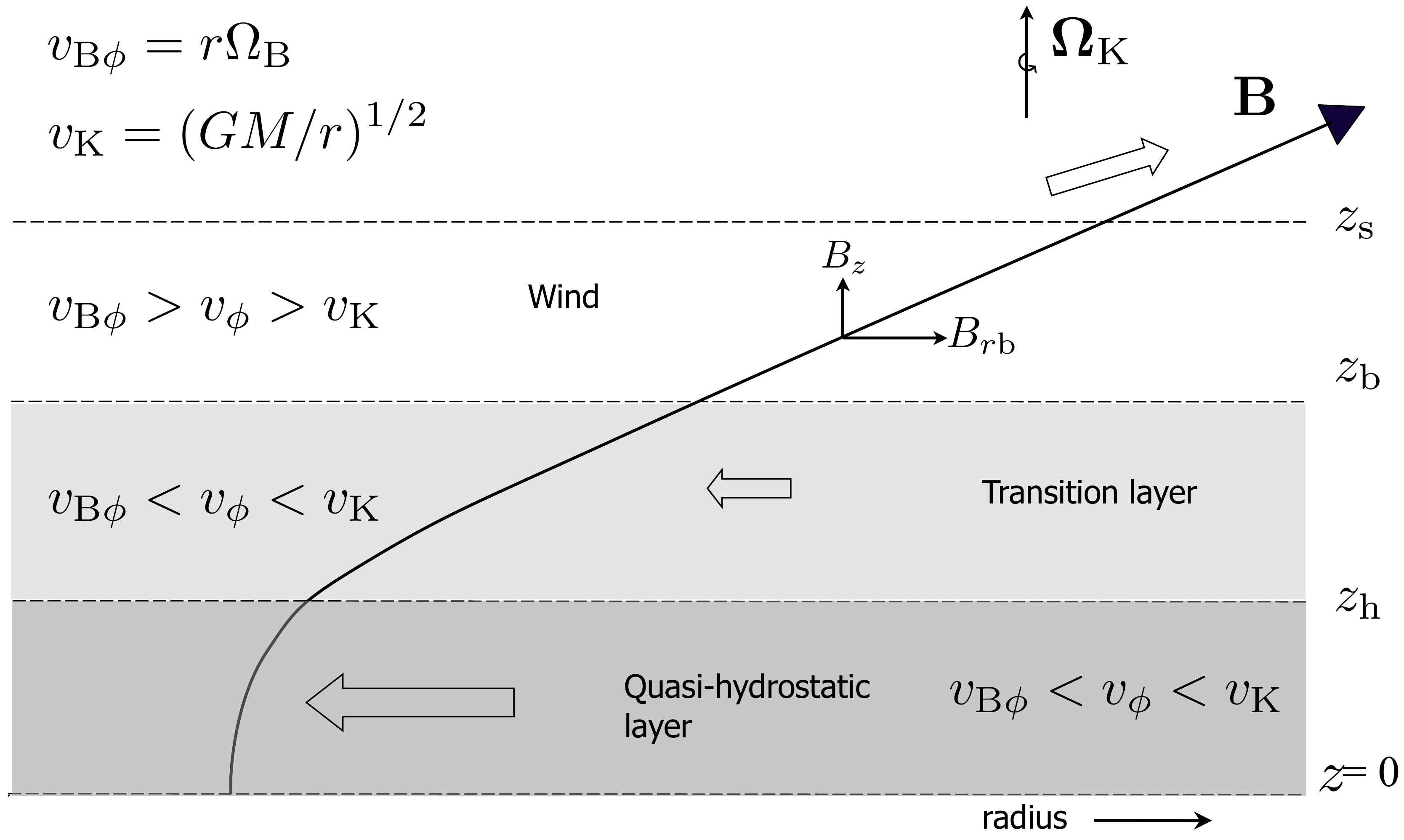}
\caption{Vertical structure of a strongly-magnetised wind-driving disc (see WK93 and also \citealt{KS11}). Three distinct layers can be identified (see text): (1) a \emph{quasi-hydrostatic layer}, next to the midplane ($z < z_{\rm h}$, where $z_{\rm h} \equiv h$ is the magnetically-reduced density scale-height), which contains the bulk of the matter; (2) a \emph{transition layer} ($z_{\rm h} < z < z_{\rm b}$, where subscript `b' denotes the launching height or base of the wind), where the magnetic energy becomes dominant and the inflow gradually diminishes; and (3) an outflow region ($z > z_{\rm b}$), which constitutes the base of the large-scale {\em wind}. $z_{\rm s}$ is the height of the sonic point and $v_{\rm B}$ is the field-line velocity. 
}
\label{fig:sim}
\end{figure*}

Fast, collimated winds are commonly observed in accreting astrophysical systems. It is likely that these outflows transport angular momentum away from the disc that surrounds the central object and thus, play an important role in regulating accretion \citep[e.g.~see reviews by][]{KP00, POFB07, KS11}. The apparent correlation between outflow and accretion signatures  in these systems \citep[e.g.][]{CES90, CA91, HEG95}, lends support to this interpretation. The details of this inflow-outflow mechanism are far from being
well understood, but it is thought that this process may be mediated by magnetic stresses [\citealt{BP82}; \citealt{WK93} (hereafter BP82 and WK93, respectively); \citealt{L96}].

Fig.~\ref{fig:illus} shows a typical disc-wind solution, plotted as a function of $z/h_{\rm T}$. This solution has been computed using the procedure detailed in KSW10 and  \citet[][hereafter SKW11]{SKW11}; in turn a generalisation of the formulation by WK93. The model parameters are:

\begin{enumerate}
\item  $a_0 \equiv v_{{\rm A}0}/c_{\rm s}$, the midplane (subscript `0') ratio of the Alfv\'en
speed to the isothermal sound speed. This parameter measures the magnetic field strength; 
\item The midplane ratios of the diffusivity tensor components: ($\eta_{\rm P}/\eta_{\perp})_0$ [or, equivalently,  ($\eta_{\rm H}/\eta_{\perp})_0$], and  ($\eta_{\perp}/\eta_{\rm O})_0$, where $\eta_{\perp} = \sqrt{\eta_{\rm P}^2 + \eta_{\rm H}^2}$ is the total diffusivity perpendicular to the magnetic field;
\item $\Lambda_{\rm 0} \equiv v_{{\rm A}0}^2/\eta_{\perp 0} \Omega_{\rm K}$, the midplane value of the {\em Elsasser number}, which measures the degree of coupling between the neutral fluid and the magnetic field (with $\Lambda \gg 1$ and $\ll 1$ corresponding to strong and weak field-matter coupling, respectively); 
\item $\epsilon \equiv -v_{r0}/c_{\rm s}$, the normalised inward radial
speed of the fluid at the midplane. This parameter is evaluated by imposing the Alfv\'en critical-point constraint on the wind solution (see SKW11);
\item $c_{\rm s}/v_{\rm K} = h_{\rm T}/r$, the ratio of the disc tidal density scale-height to the radius (e.g.~the disc geometric thickness); and  
\item $\epsilon_B \equiv -v_{\rm Br0}/c_{\rm s} = -cE_{\phi 0}/c_{\rm s} B_z$, the normalised (and constant with $z$) azimuthal
component of the electric field\footnote{\label{ansatz}Setting 
$\epsilon_{\rm B} = 0$ effectively fixes the value of $B_r$ at the disc surface. In a 
global formulation, $B_r$ would be determined by the radial distribution of $B_z$ \citep[e.g.][]{OL01, KK02}.}. This parameter measures the radial drift speed of the poloidal magnetic field lines  (WK93). 
\end{enumerate}
We will also use the expression  $\Upsilon \equiv \rho_{\rm i} \gamma_{\rm i}/\Omega_{\rm K}$, the ratio of the Keplerian rotation time to the neutral-ion momentum exchange time (e.g.~WK93; see also footnote \ref{foot:note2}). It can be shown that in the ambipolar regime  $\Lambda = \Upsilon$, whereas in the Hall limit $\Lambda = \Upsilon |\beta_{\rm i}|$   (e.g.~KSW10).

The left panel of Fig.~\ref{fig:illus} displays the gas density (normalised by its value at the midplane) as well as the radial and azimuthal components of the magnetic field (normalised by the vertical field, which under the adopted thin-disc approximation is vertically constant), as a function of height above the midplane, expressed in units of the tidal scale-height. The right panel shows corresponding runs for all velocity components, measured with respect to the Keplerian velocity $v_{\rm K}$ and normalised by the isothermal sound speed $c_{\rm s}$. 
This solution illustrates the main features of winds accelerated centrifugally from strongly-magnetised discs (WK93). Three distinct layers can typically be identified.  These layers are presented schematically in Fig.~\ref{fig:sim}, and described below. 

\begin{itemize}
\item The \emph{quasi--hydrostatic} layer, adjacent to the disc midplane ($z < z_{\rm h} \equiv h$, where $h$ is the magnetically-reduced density scale-height\footnote{Magnetic compression dominates over the gravitational tidal squeezing of the disc in all viable wind-driving disc solutions of interest here (WK93, KSW10), so that the magnetically-compressed density scale-height ($h$) satisfies $h/h_{\rm T} < 1$.}), is matter dominated ($v_{\rm B \phi} < v_\phi < v_{\rm K}$, where $v_{\rm B}$ is the field-line velocity). 
In this layer, the topology of the magnetic field is consistent with the field lines being radially bent and azimuthally sheared. The field removes angular momentum from the gas, as a result of the ion-neutral drag, and matter moves radially inwards. 
\item The \emph{transition} region, above the quasi--hydrostatic layer ($z_{\rm h} < z < z_{\rm b}$, where subscript `b' denotes the base of the wind), where magnetic energy comes to dominate over the thermal energy of the fluid as a result of the strong drop in density away from the midplane. The magnetic field lines are locally straight in this zone (note that they are still inclined outwards). Here (as in the quasi-hydrostatic layer at lower $z$), the fluid azimuthal velocity is sub-Keplerian, as magnetic tension helps support the fluid against the gravitational pull of the central protostar. 
\item As the angular velocity of the field lines is constant, their azimuthal velocity increases along the line, until they eventually overtake the fluid in which they are embedded. This location marks the beginning of the uppermost \emph{wind} region ($z > z_{\rm b}$), which constitutes the base of the outflow. In this layer the field transfers angular momentum back to the matter, accelerating it centrifugally; the radial and vertical velocity components are positive, and the azimuthal velocity is super-Keplerian ($v_{\rm B \phi} > v_\phi > v_{\rm K}$).
\end{itemize}

In the specific solution shown in Fig.~\ref{fig:illus}, the quasi-hydrostatic layer, which encompasses the section of the disc associated with strong gradients in the fluid density and magnetic field components, extends from $z/h_{\rm T} = 0$ to $z/h_{\rm T} \approx 0.4$ (the region shaded in red in the left panel of the figure). The transition region lies directly above it, and extends up to the height where the azimuthal velocity becomes Keplerian (up to the vertical red line in the right panel), so it constitutes the region where $0.4 \lesssim z/h_{\rm T} \lesssim 1.8$. Finally, the wind region satisfies $z/h_{\rm T} > 1.8$. Note that in the first two layers (i.e.~within the {\em disc}) the radial velocity is negative, the azimuthal velocity is sub-Keplerian and the vertical velocity is small. In the {\em wind} layer, on the other hand, all these velocities [$(\mathbf{v} - v_{\rm K}\hat{\mathbf{\phi}})/c_{\rm s}$] are positive and increasing with $z$.
 
\begin{figure*}
\centering
\includegraphics[scale=1.2]{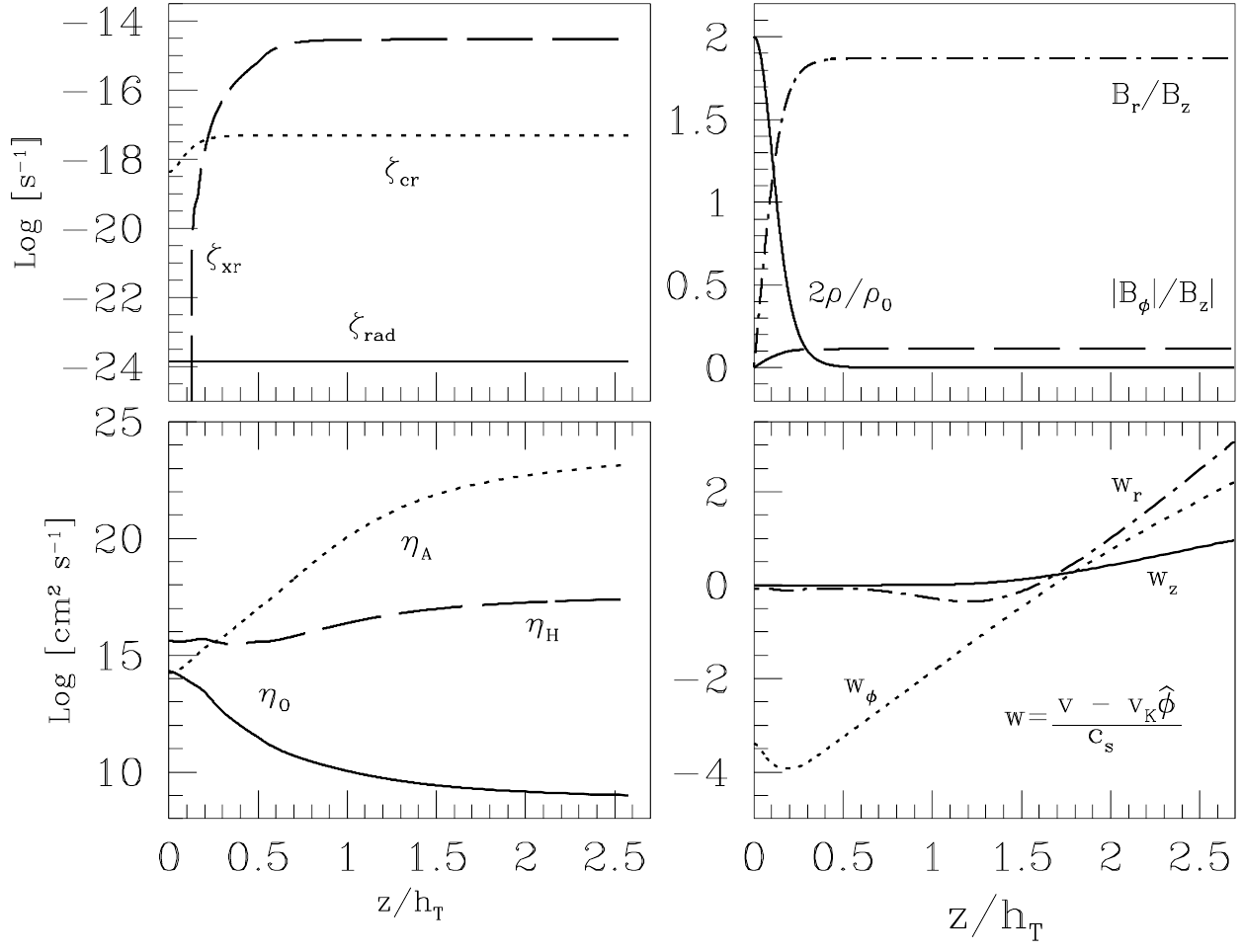}
\caption{Representative radially-localised disc-wind solution at $r = 1$ AU, matched to the global, self-similar models of BP82, for a disc surrounding a solar-mass protostar. The surface density of the disc is $\Sigma = 600$ g cm$^{-2}$ and the remaining free parameters are: $a_{\rm 0} = 0.75$, $v_{\rm K}/c_{\rm s} = 10$, $\epsilon = 0.1$ and $\epsilon_{\rm B} = 0$.   The magnetic diffusivity components and field-matter coupling (measured by the Elsasser number) are calculated self-consistently by evaluating the ionisation balance in the disc (see text).   The parameters of the self-similar model are $\kappa = 2.6 \times 10^{-6}$, $\lambda = 4.4 \times 10^{3}$, and $\xi'_{\rm b} \equiv B_{r{\rm b}}/B_z = 1.6$. {\em Left column}. Top panel: Ionisation rates contributed by cosmic rays (cr), X-rays (xr) and radioactive decay (rad). Bottom panel: Ambipolar, Hall and Ohm diffusivity components.  {\em Right column}. Top panel: Density and magnetic field components in the plane of the disc (radial and azimuthal). Bottom panel: Velocity components. The resulting local mass accretion rate of the model is $7 \times 10^{-6} \ M_\odot \ {\rm yr}^{-1}$, a value that is consistent with the accretion rates inferred for objects in the early stages of the protostellar accretion phase (e.g.~Class 0/Class I objects; e.g.~\citealt{HEG95}).
}
\label{fig:real_sol}
\end{figure*}

Solutions such as the one shown in Fig.~\ref{fig:illus} are found for relatively strong magnetic fields, such that $a_{\rm 0} \lesssim 1$. Under these fluid conditions, the MRI is expected to be suppressed because the wavelength of the critical 
MRI-unstable mode (longward of which the instability operates; see \citealt{BH91}) exceeds the magnetically-reduced density scale-height $h$ (WK93).  For the particular solution shown in the figure, the magnetic field is strongly coupled to matter even at the disc midplane ($\Lambda_{\rm 0} \gg 1$), and ambipolar diffusion dominates over the entire cross-section of the disc. For simplicity, the diffusivity components are assumed to scale with height in such a way that the local Elsasser number ($\Lambda$) remains constant with $z$. Under these approximations, it is self-consistent to find relatively high values of the inward velocity of the fluid at the disc midplane (of the order of the isothermal sound speed, such that the parameter $\epsilon$ is in the range of 0.3 -- 1). A more realistic treatment of the fluid conditions would incorporate the vertical stratification of the diffusivity, with different regimes dominating at different vertical locations (see section \ref{subsec:Magdiff}), and possibly a magnetically weakly-coupled region in the disc interior. In the latter case, in particular, more moderate values of the inward flow speed at the midplane are expected \citep[][see also Fig.~\ref{fig:real_sol} below]{L96, W97}.

The next issue to consider is that in order to compute self-consistent, coupled disc-wind solutions, it is essential to solve simultaneously for the complex interactions between the gas and the magnetic field within the disc -- where the launching process occurs -- as well as to follow the acceleration of the wind past the critical surfaces of the flow. Physical quantities of critical importance, such as the gas density and the ionisation fraction, are expected to vary by many orders of magnitude between the disc midplane and its surface. For example, the gas density typically drops by $\sim 6$ orders of magnitude across the half-thickness of the disc. 
As a result, global simulations of these winds have adopted a number of simplifications to treat the underlying disc. In some (including the pioneering work of BP82),  the disc is treated as a boundary condition, outside of the computational domain. Other authors have attempted to incorporate the disc, either via assuming perfect field-matter coupling (i.e.~negligible diffusivity)  
or through the prescription of a simplified 
finite resistivity  term 
(e.g.~see the review by \citealt{POFB07} and also \citealt{FP95, L96, CF00, Zanni07a}). This approach is appropriate to follow the propagation of the outflow to large distances for an essentially \emph{prescribed} wind mass flux, but it can not be used to determine, self-consistently, whether the wind is \emph{launched} in the first place.

A complementary approach, first developed by WK93 for an ambipolar-diffusion-dominated disc, is to resolve explicitly the vertical stratification of the fluid variables, ionisation fraction and magnetic diffusivity, as well as the resulting field-matter coupling, for a radially-localised region of the disc. This formulation systematically reduces the nonlinear governing equations that control the evolution of the fluid into a set of ordinary differential equations (ODE) in $z$. This system describes, for a particular radial location, the vertical structure of the disc -- and the base of the outflow -- from the midplane up to the critical (sonic) surface of the flow (see also KSW10 for a generalisation of this procedure to a tensor diffusivity). 
The system of ODE is numerically integrated by applying boundary conditions at the midplane and at the sonic point (see WK93 and SKW11 for details of the numerical procedure). The ionisation balance is calculated by evaluating the action of ionisation processes (driven by cosmic rays, X-rays emitted by the central object and  
radioactive decay) and recombinations occurring, in general, both in the gas phase and on grain surfaces (see Section \ref{subsec:Magdiff}). The X-ray ionisation rate is taken from the Monte Carlo calculations of \citet{IG99}, whereas that of cosmic rays  is calculated by attenuating the standard (``canonical") rate in the interstellar medium as the cosmic rays penetrate the disc [the canonical rate is taken to be $10^{17}$ (s$^{-1}$) per hydrogen atom; see e.g.~\citealt{UN81}]. 
The field-matter diffusivity is treated as a tensor, incorporating the Hall, Ohmic and Ambipolar terms (Section \ref{subsec:Magdiff}). This approach eliminates the need to keep separate equations of motion for each fluid component and greatly simplifies the calculations, particularly when dust grains are present. 

 \begin{figure*}
\centering
\includegraphics[scale=0.61]{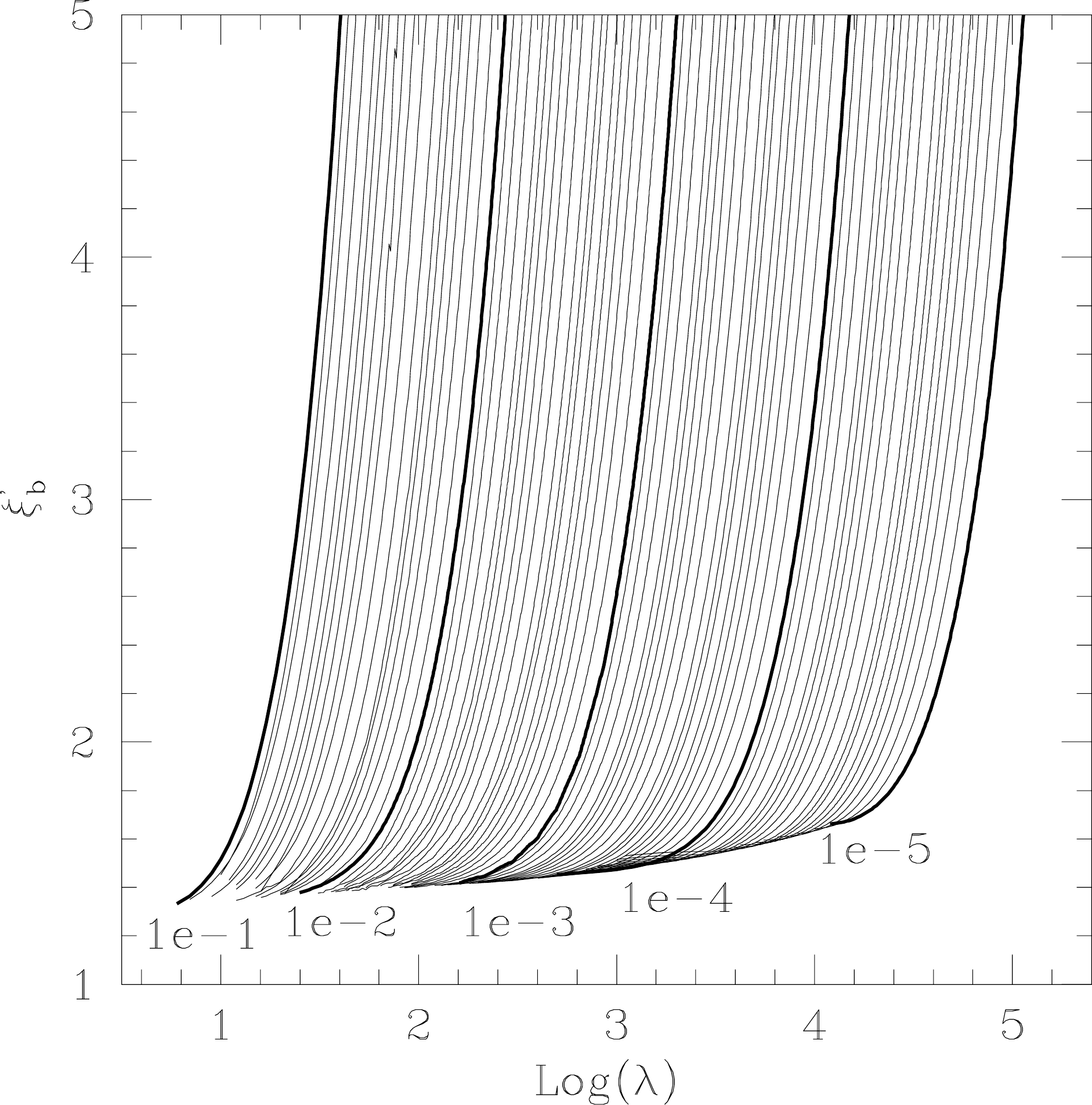}
\caption{Global, self-similar wind solutions of BP82 in parameter space. The solutions are plotted in a $\xi_{\rm b}'$ (= $B_{\rm rb}/B_z$) -- log $\lambda$  plane, where $\lambda$ is the normalised total specific angular momentum carried away by the wind (including the matter and the magnetic field contributions) and  $\xi_{\rm b}'$ measures the inclination of the field lines at the base of the outflow. The curves are labelled by the normalised mass-to-magnetic flux ratio ($\kappa$) and the values of $\kappa$ that correspond to the darker curves are indicated in the figure. See SKW11 for the method of calculation of these solutions and for a tabular compilation of the results. 
}
\label{fig:windres}
\end{figure*}

The methodology just described is appropriate to model the launch of the wind, but it cannot be used to follow the propagation of the outflow far from the source, where the thin-disc approximation breaks down. It is, however, clearly necessary 
to ensure that the obtained `disc solution' continues to accelerate past the sonic point. To ensure that this is the case, we evaluate the parameters of the outflow at the disc surface and match this local solution to a global wind model (we use for this purpose the BP82 self-similar wind solutions). In order to facilitate the matching, we constructed a library of these self-similar `wind solutions' for a wide range of their model parameters, which are:  the normalised mass-to-magnetic flux ratio ($\kappa$), the normalised total specific angular momentum carried away by the wind ($\lambda$, which incorporates contributions from the matter and from the magnetic field), and the inclination of the field lines at the disc surface ($\xi_{\rm b}' =  B_{\rm rb}/B_z$). The combinations of these parameters corresponding to viable global wind solutions are shown graphically in Fig.~\ref{fig:windres}. The complete set of these results is also available in the electronic version of SKW11. The described overall methodology is useful to explore the viability of these winds as a function of radius from the central object, as well as to obtain a realistic estimate of the wind mass loss rate, which could be used to constrain theoretical models against observational results.

A representative local solution, matched to the BP82 global, self-similar models, is shown in Fig.~\ref{fig:real_sol} for $r = 1$ AU in a disc surrounding a solar-mass protostar. The surface density is $\Sigma = 600$ g cm$^{-2}$ and the remaining free model parameters are: $a_{\rm 0} = 0.75$, $v_{\rm K}/c_{\rm s} = 10$, $\epsilon = 0.1$ and $\epsilon_{\rm B} = 0$. The magnetic diffusivity components and field-matter coupling ($\Lambda$) are calculated self-consistently by evaluating the ionisation balance of the fluid (see Section \ref{subsec:Magdiff}). On the left-hand side of the figure, the top panel shows the ionisation rates contributed by cosmic rays (curve labelled by the subscript `cr'), X-rays (`xr') and radioactive decay (`rad'); and the bottom panel displays the resulting diffusivity terms. On the right-hand side, the normalised density and magnetic field components in the plane of the disc are shown in the top panel. The normalised velocity components are displayed in the bottom panel. The global wind model parameters are: $\kappa = 2.6 \times 10^{-6}$, $\lambda = 4.4 \times 10^{3}$, and $\xi'_{\rm b} \equiv B_{r{\rm b}}/B_z = 1.6$.  The local mass accretion rate for this model is $\sim 7 \times 10^{-6} \ M_\odot \ {\rm yr}^{-1}$, which is consistent with the inferred values for the early (Class 0/Class I) protostellar accretion phase \citep[e.g.][]{HEG95}. 

We have also constrained the parameter space occupied by physically-viable wind-driving disc solutions (see WK93 for the original derivations in the ambipolar diffusion limit; as well as KSW10 for the generalisation to a tensor diffusivity and application to the Hall and Ohm regimes).  Using the hydrostatic approximation (e.g. neglecting the vertical velocity component of the fluid), we are able to show that the solutions are required to satisfy the following constraints.
\begin{itemize}
\item {\em Sub-Keplerian flow below the launching region}. Below the launching height the azimuthal velocity of the gas should be sub-Keplerian, as discussed earlier in this section. 
\item {\em Wind-launching criterion}. The sufficient inclination of the field lines with respect to the rotation axis of the disc for centrifugal acceleration to occur (BP82). 
\item {\em Location of the base of the wind}. Both theoretical and observational arguments suggest that only the uppermost section of the disc should participate in the outflow \citep{KP00}. We apply this condition by requiring that the base of the wind be located above one magnetically-reduced density scale-height, or $z_{\rm b}/h > 1$. It can be shown that when this condition is violated, the azimuthal component of the magnetic field changes sign within the disc, indicating an attempt by the field to transfer angular momentum back to matter below the wind launching height.
\item {\em Dissipation rate}. The rate of heating by Joule (e.g. magnetic) dissipation at the disc midplane is restricted to be less than the rate of gravitational potential energy release at that location \citep{K97}. 
\end{itemize}

\setlength{\cellspacetoplimit}{1mm}
\setlength{\cellspacebottomlimit}{1mm}
\begin{table*}
\caption{Parameter constraints for wind-driving disc solutions in the
limit where the Hall diffusivity dominates. Four distinct cases can be identified, depending on 
whether the combinations $\beta_{\rm e 0} \beta_{\rm i 0}$ and  $2\Lambda_0 \ (=
2\Upsilon_0 |\beta_{\rm i 0}|)$ 
are larger or smaller than unity (KSW10). The first inequality
expresses the requirement that the disc remain sub-Keplerian below the
launching point, the second is the wind
launching condition (the requirement that the magnetic field lines be
sufficiently inclined with respect to the angular velocity vector of the disc for centrifugal acceleration to
occur), the third ensures that the base of the wind is located above the
magnetically-reduced density scale-height, and the fourth specifies
that the rate of Joule heating at the midplane should not exceed the
rate of release of gravitational potential energy at that location.}
\label{table:constraints} 
\begin{center}
\begin{tabular}{ScScScScScScScScScScScSc} 
\hline 
{\normalsize Case} & \multicolumn{2}{c}{{\normalsize Limits}} &
\multicolumn{9}{c}{{\normalsize Parameter Constraints -- Hall Limit }}\\ 
& \  $\beta_{\rm e 0} \beta_{\rm i 0}$ & \ \ \ $\Lambda_0=\Upsilon_0 |\beta_{\rm i
0}|$ & \multicolumn{9}{c}{{\small 
}}\\ 
\hline 
(i) & $> 1$ & $> 1/2$ & \ \ \  $(2\Upsilon_0)^{-1/2}$ & $\lesssim$ & $a_0$ &
$\lesssim$ & $2$ & $\lesssim$ & $\epsilon\Upsilon_0$ & $\lesssim$ & $\vk/2
c_{\rm s}$ \\ 

(ii) & $ > 1$ & $< 1/2$ & \ \ \ $\betio^{1/2}$ & $ \lesssim$ &
$a_0$ & $\lesssim$ & $2 (\Upsilon_0 \betio)^{1/2}$ & $\lesssim$ & $\epsilon/
2\betio$ & $\lesssim$ & $\Upsilon_0 \betio \vk/ c_{\rm s}$ \\ 

(iii) & $< 1$ & $>
1/2$ & \ \ \  $(2\Upsilon_0)^{-1/2}$ & $\lesssim$ & $a_0$ & $\lesssim$ & $2$ &
$\lesssim$ & $\epsilon \Upsilon_0 \beta_{\rm e0} \beta_{\rm i0}$ & $\lesssim$
& $\vk/2 c_{\rm s}$ \\ 

(iv) & $< 1$ & $< 1/2$ & \ \ \  $\betio^{1/2}$ & $\lesssim$ & $a_0$ & $\lesssim$
& $2(\Upsilon_0 \betio)^{1/2}$ & $\lesssim$ & $\epsilon \beta_{\rm e0}/2$ &
$\lesssim$ & $\Upsilon_0 \betio \vk/ c_{\rm s}$ \\ 
\hline 
\end{tabular} 
\end{center} 
\end{table*}

The above constraints can be shown to lead to four parameter sub-regimes where viable solutions exist in the Hall regime, and to three in the Ohm regime (I refer the reader to KSW10 for more details). The constraints for the Hall regime are summarised in Table \ref{table:constraints}. These are differentiated by the values of the combinations $\beta_{\rm e 0} \beta_{\rm i 0}$ and  $2\Lambda_{\rm 0}$ ($=
2\Upsilon_{\rm 0} |\beta_{\rm i 0}|$ in this limit) in comparison with unity.  Our numerical solutions (SKW11) confirm these predictions: We find that no physically-relevant solutions are found outside the boundaries specified by these constraints. 

We also explored the dependence of the solutions on the model parameters. We  found that increasing the relative contribution of the Hall diffusivity results in a smaller magnetically-reduced density scale-height, and in a higher location above the disc midplane of both the base of the wind and the sonic surface. As a result, the density at the sonic point, and the associated mass outflow rate, are reduced. 
Our calculations also show that, in all diffusivity regimes, viable solutions satisfy the requirement that the neutral -- ion momentum exchange time be shorter than the disc orbital time ($\Upsilon \gtrsim 1$). This is, therefore, a fundamental constraint on the wind solutions considered here.

\begin{figure}
   \centering
   \includegraphics[width=0.48\textwidth]{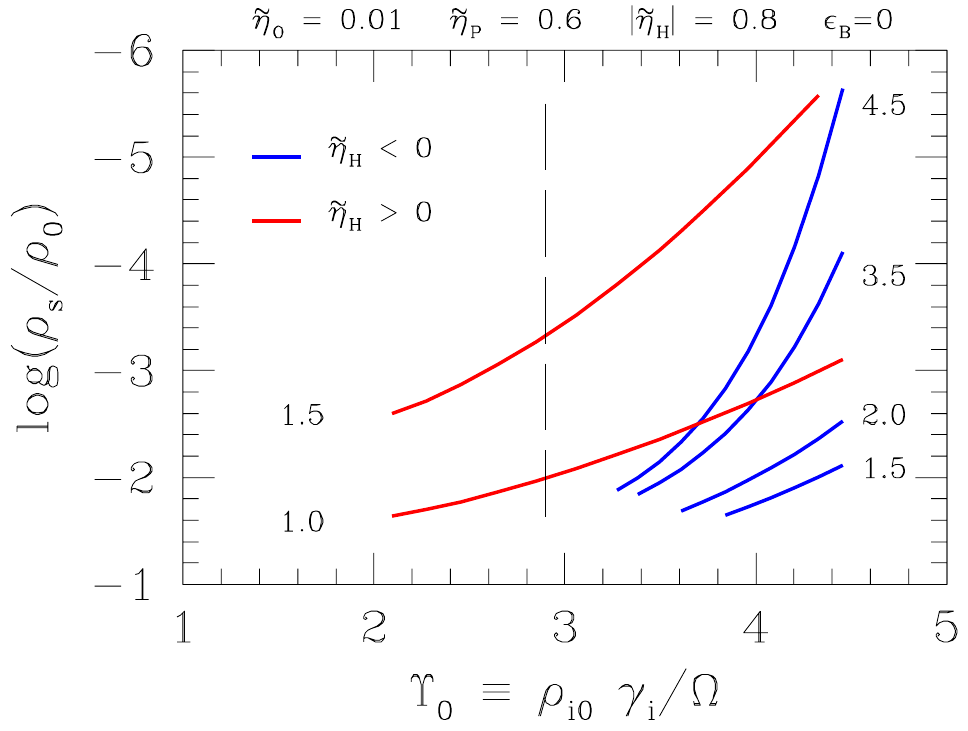} 
\caption{Normalised sonic-point density ($\rho_{\rm s}/\rho_{\rm 0}$) as a
function of the parameter $\upo$ 
for solutions in the Hall sub-regime~(i) (KSW10, see also Table \ref{table:constraints}). These results illustrate the dependence of physically-viable wind-driving disc models on the orientation of the magnetic field when the Hall diffusivity dominates. 
The red curves correspond to $\tilde{\eta}_{\rm H}   \equiv \eta_{\rm H} \Omega/ c_{\rm s}^2 > 0$ (or $B_z > 0$), and the blue ones  to $\tilde{\eta}_{\rm H} < 0$ ($B_z < 0$). These are labelled by the normalized midplane inflow velocity $\epsilon$. The diffusivity values listed at the top of the figure refer to the disc midplane.
The value of $\betiabs^{-1}$ is the same ($1.45$) for all the
solutions. The results are broadly consistent with the prediction of the
hydrostatic analysis (KSW10) that there should be no viable solutions for $\tilde{\eta}_{\rm H} < 0$ when $\upo$ drops below $2 \betiabs^{-1}$ (i.e.~to the left of the vertical dashed line). Where solutions with both polarities exist, their properties are different, as predicted. 
}
\label{fig:5_26}
\end{figure}

Finally, we found that the magnetic field polarity modifies both the properties of the solutions and the parameter ranges where these exist when the Hall regime is dynamically important. This reflects the dependence of the Hall diffusivity on the sign of $B_z$. Specifically, our analysis shows that for two of the Hall sub-regimes (Cases $i$ and $iii$ in Table \ref{table:constraints}), positive and negative polarity solutions have different properties:  For the same values of all the other parameters, both the base of the wind and the sonic point are located higher above the midplane when the magnetic field is aligned with the angular velocity vector of the disc ($B_z > 0$) than when it is pointing in the other direction ($B_z < 0$). The mass outflow rate is correspondingly lower in the former case than in the latter. Furthermore, no viable solutions are predicted to exist in these sub-regimes when $\upo < 2 \betiabs^{-1}$ and $B_z < 0$. For the other two parameter sub-regimes (Cases $ii$ and $iv$ in Table \ref{table:constraints}), the situation is even more drastic: Solutions are predicted to exist for the positive polarity only (e.g.~no viable solutions are expected for the negative polarity). 

Our numerical solutions confirm the above expectations. This is illustrated in Fig.~\ref{fig:5_26} (for sub-regime $i$). This figure shows the normalised density at the sonic point ($\rho_{\rm s}/\rho_{\rm 0}$) as a function of the parameter $\upo$, for solutions with both (positive and negative) magnetic field polarities. In all cases, the value of $\betiabs^{-1} = 1.45$. The displayed results confirm the prediction of the hydrostatic analysis (KSW10) that no viable solutions exist for $\tilde{\eta}_{\rm H} \equiv \eta_{\rm H} \Omega/ c_{\rm s}^2 < 0$ when $\upo < 2 \betiabs^{-1}$ (i.e.~to the left of the vertical dashed line). It is also evident that in the region of parameter space where solutions with both polarities exist, these have different properties. Note that the mass outflow rate (measured by $\rho_{\rm s}/\rho_{\rm 0}$) is smaller in the $B_z > 0$ case,  in accordance with our predictions.

\section{Radial transport via MRI-induced turbulence}
\label{sec:MRI}

  \begin{figure*}[]
\centering
\includegraphics[width=6.3in]{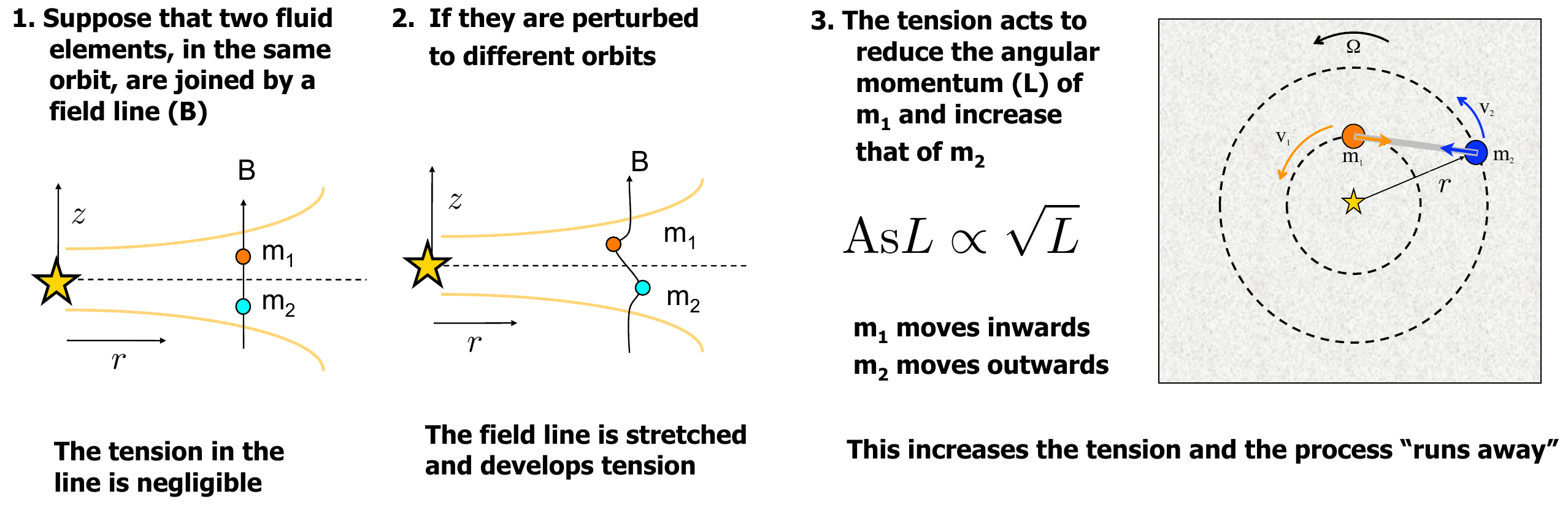}
\caption{Physical principle underpinning the action of the magnetorotational instability (MRI). The left panel shows a schematic diagram (in the $r$-$z$ plane) of a protostellar disc threaded by a vertical magnetic field. Two fluid elements are located at the same orbital radius, joined by the field line. This panel corresponds to the initial configuration of the disc. Once perturbations are initiated (middle panel), these will amplify (right panel, showing the disc viewed from the top). See text for details.}
\label{fig:MRIcartoon}
\end{figure*}

The MRI (\citealt{BH91}, see also the review by \citealt{BH98}), transfers angular momentum radially outwards via magnetic field lines that connect fluid elements located at different radii. The physical principle that underpins this mechanism is illustrated in Fig.~\ref{fig:MRIcartoon}. Imagine two fluid elements that are initially located at the same orbital radius and are joined by a vertical magnetic field line threading the disc (left panel). In this configuration the magnetic tension in the line is negligible. Now imagine that the elements are perturbed from their initial orbits, so that one of them (the one labelled $m_{\rm 1}$ in the figure) is displaced to an inner orbit, whereas the other ($m_{\rm 2}$) moves to an outer orbit (middle panel). Assuming that the field-matter coupling is sufficient for the magnetic field line to be dragged by the moving fluid elements, the line will be stretched  and magnetic tension will develop. Note that the element in the inner orbit ($m_{\rm 1}$) overtakes the one in the outer orbit ($m_{\rm 2}$), as a result of the dependence of the Keplerian orbital motion with radius ($v_{\rm K} = \sqrt{GM/r}$). The tension in the line, therefore, acts to reduce the angular momentum of the inner element, and to increase that of the outer element (right panel). As a result of this process, angular momentum is transferred from $m_{\rm 1}$ to $m_{\rm 2}$, causing the former element to move further inwards while the latter retreats to an even larger radius. As the elements continue to separate, the tension in the magnetic field line joining them increases, and the process ``runs away"; perturbations, once initiated, will amplify. 

The nonlinear stages of the instability have been studied extensively \citep[e.g. see the reviews by][]{balbus03a, BH98}.  The result is MHD-driven turbulence where angular momentum is transported radially outwards. This outward transport is linked to the fact that the turbulence is anisotropic:  the radial and azimuthal components of the perturbations of the magnetic and velocity fields are significantly correlated. In other words, the transport of angular momentum follows directly from the statistical correlation in the Reynolds (velocity) and Maxwell (magnetic field) stress tensors \citep{BH98}.

Under ideal-MHD conditions (negligible magnetic diffusivity), MRI-unstable modes grow in weakly-magnetised discs (meaning here that the magnetic energy density is small in comparison to the thermal energy density; or $a \equiv v_{\rm A}/c_{\rm s} \ll 1$), provided that the angular velocity of the fluid decreases radially outwards. This condition is satisfied in Keplerian discs, for which $\Omega_{\rm K} = v_{\rm K}/r \propto r^{-3/2}$. An analysis of the dispersion relation, linking the wavenumbers and growth rates of the MRI-unstable modes \citep[e.g.][]{BH91}, shows that the maximum growth rate of the instability is of the order of $\Omega_{\rm K}$, and it is independent of the field strength. 

The upper limit on the range of magnetic field strengths for which the MRI is expected to grow in discs (under ideal-MHD conditions) reflects the fact that the critical wavelength (longward of which the instability operates) is directly proportional to $B$ \citep{BH91}. For a sufficiently strong field, therefore, the critical wavelength will exceed the thickness of the disc ($\sim h_{\rm T}$) and the instability will be damped. It is also the case that the wavelength of the most-unstable mode is larger than the critical wavelength \citep{BH91}. Therefore, even if the wavelength of maximum growth does not fit within the disc, there may be slower-growing modes that are able to develop (provided that the disc thickness is larger than the critical wavelength of the instability). In these conditions, MRI-induced turbulence is still viable, although at a reduced  strength. Because of the general conditions under which it operates, the MRI is a promising mechanism for the redistribution of angular momentum in astrophysical accreting systems.
 \begin{figure*}[]
\centering
\includegraphics[width=4.5in]{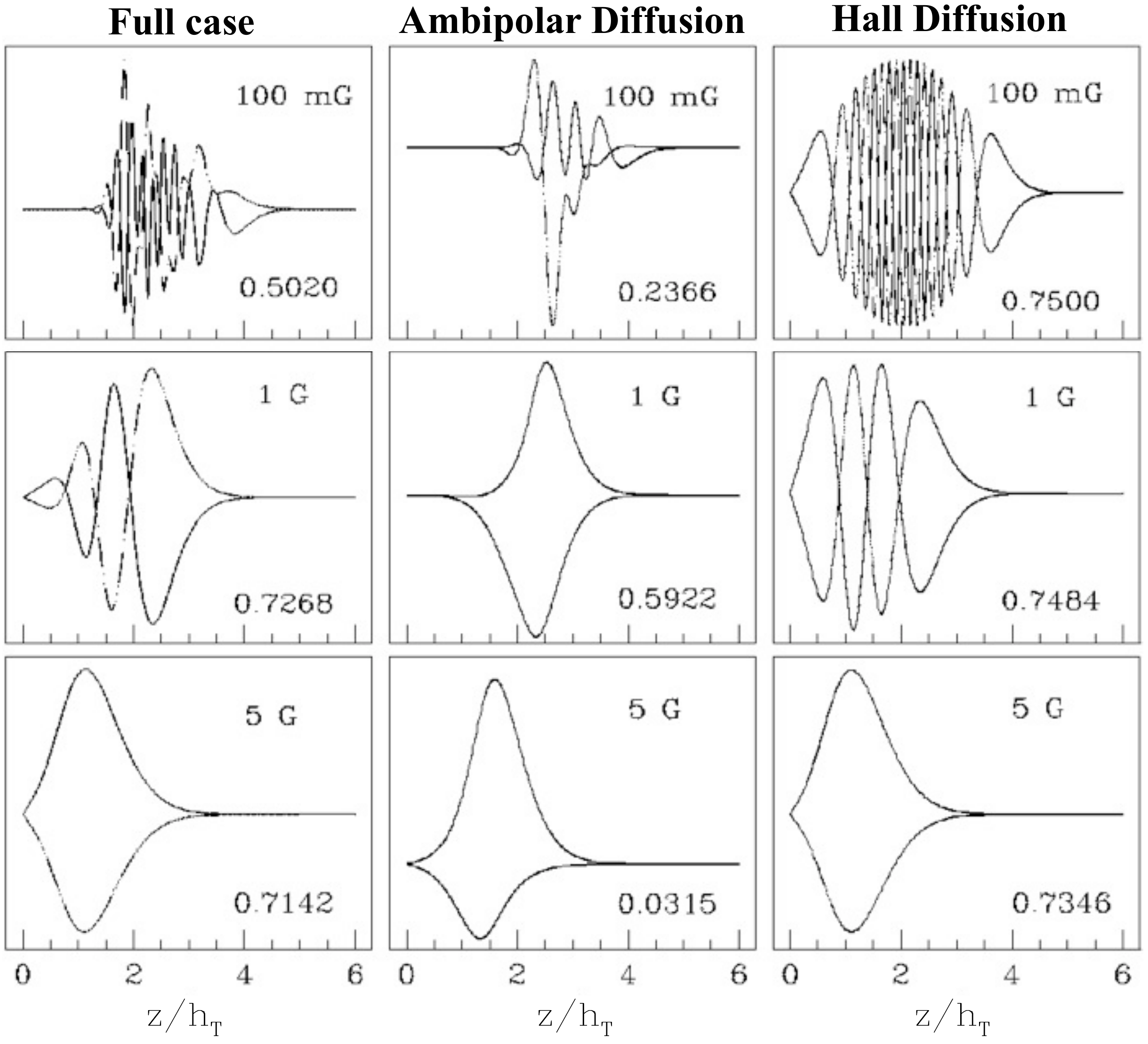}
\caption{Vertical structure and linear growth rate of the most-unstable MRI modes for a minimum--mass solar nebula disc at $r = 1$ AU, as a function of the magnetic field strength (shown in the upper-right corner of each panel) and for different diffusivity limits. In each panel, the solid (dashed) line denotes the radial (azimuthal) perturbation of the magnetic field, which is initially vertical. The growth rate (in units of the Keplerian frequency $\Omega_{\rm K}$) is indicated in the lower-right corner of each panel. The middle and right columns show solutions obtained under the Ambipolar and Hall diffusion ($B_z > 0$) approximations, respectively. The left column depicts solutions incorporating both diffusivity components. These results show that MRI perturbations grow faster, and are active closer to the midplane, when both diffusion mechanisms are considered.
}
\label{fig:MRI1}
\end{figure*}

 \begin{figure}
\centering
\includegraphics[width=3.0in]{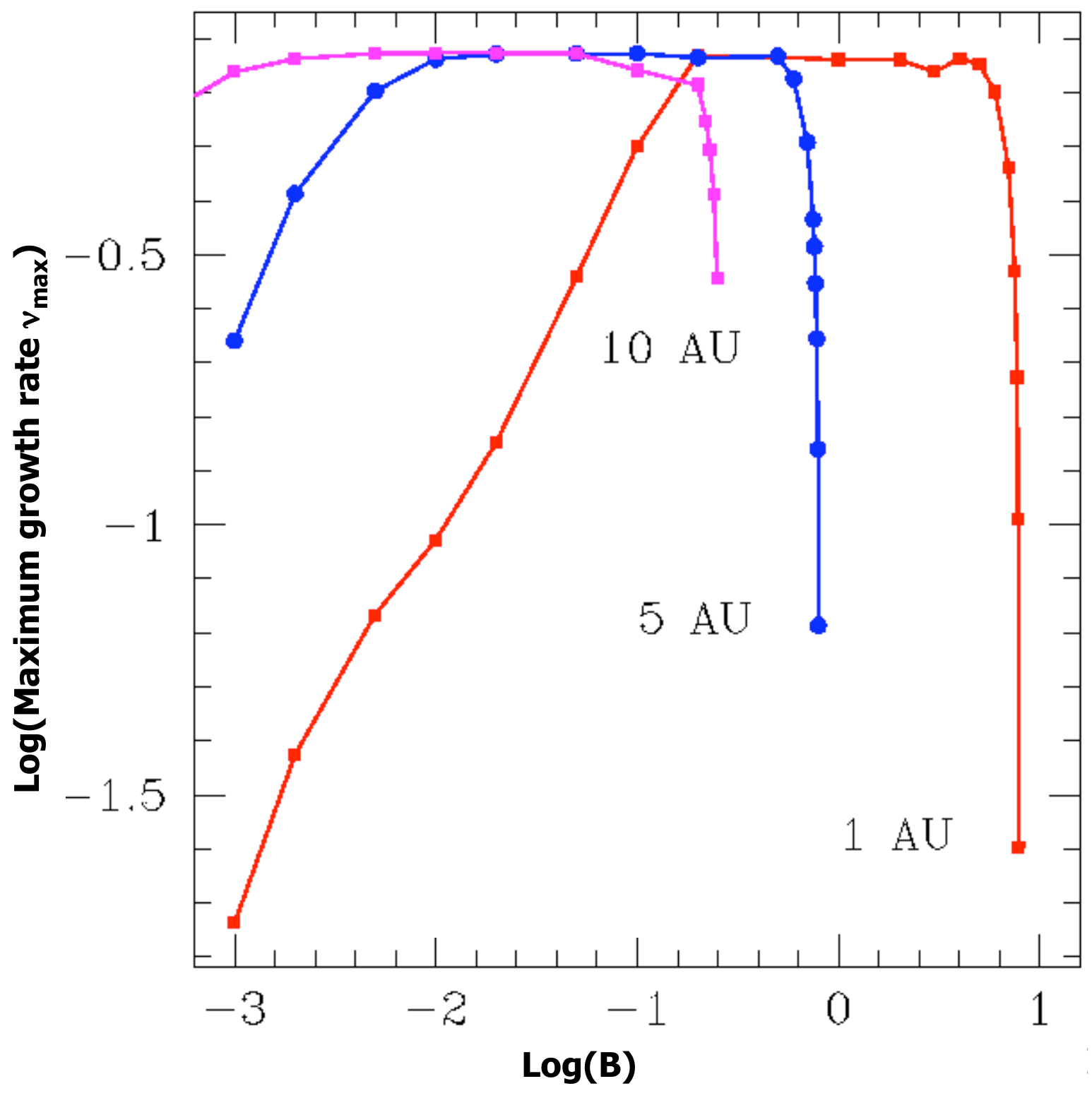}
\caption{Linear growth of the most-unstable MRI modes for a minimum--mass solar nebula model at $r = 1$, $5$ and $10$ AU as a function of the strength of the magnetic field. For each radius, the maximum field strength for which the instability grows is the one for which the ratio of the Alfv\'en speed to the isothermal sound speed at the midplane is of order unity ($a_{\rm 0} \approx 1$).  For this $B$, the critical wavelength (longward of which the MRI operates) is of the order of the disc thickness ($\sim h_{\rm T}$) and the instability is damped. }
\label{fig:MRI2}
\end{figure}

Ideal-MHD conditions, however, are not likely to be satisfied in the deep, dense interiors of protostellar discs. In these environments, therefore, it is essential to consider how the high magnetic diffusivity impacts on the development and properties of the instability. Fig.~\ref{fig:MRI1} illustrates this point. This figure shows the vertical structure and the linear growth rate of the fastest-growing MRI perturbations 
for a minimum--mass solar nebula disc at $r = 1$ AU, as a function of the magnetic field strength and for different diffusivity limits (Ambipolar diffusion, Hall diffusion with $B_z > 0$, and the `Full case', where both components of the diffusivity tensor are present). In each panel, the solid (dashed) line denotes the radial (azimuthal) perturbation of the magnetic field, which is initially vertical. The field strength is indicated in the top-right corner of each panel, whereas the growth rate, in units of $\Omega_{\rm K}$, is shown in the lower-right corner. This notation is observed here in all plots that display the structure of the instability. These solutions incorporate a realistic ionisation profile, assuming that charges are carried by ions and electrons only (e.g.~dust grains have settled out of the gas phase).  They were obtained using the procedure detailed in \citet{SW03, SW05}, to which I refer the reader for additional details. Note that Hall diffusion strongly modifies the structure and growth of the MRI-unstable modes. Specifically, when this diffusion mechanism is included in the calculations the vertical extension of the magnetically-inactive zone straddling the disc midplane is significantly reduced, and the growth rates of the perturbations increase in comparison to those obtained in the ambipolar diffusion limit. Note that for the field strengths shown, the perturbations in the Hall limit grow at about the ideal-MHD rate, whereas the ones in the ambipolar diffusion limit grow at a significantly reduced rate.  

WS11 have conducted a local linear analysis of the MRI, for a weak vertical field and considering axisymmetric perturbations with a solely vertical wavevector. This study demonstrated that for a given value of the Pedersen diffusivity ($\eta_{\rm P}$), the maximum growth rate increases with the Hall diffusivity term ($\eta_{\rm H}$). In particular, for a given $\eta_{\rm P}$ the growth rate vanishes when $\eta_{\rm H} \Omega_{\rm K}/ v_{\rm A}^2 \leq -2$, and it increases to the ideal-MHD maximum of $\sim 0.75\Omega_{\rm K}$ as $\eta_{\rm H}  \rightarrow \infty$. 
The authors also show that, for the minimum-mass solar nebula model and at $r = 1$ AU, Hall diffusion modifies the extent of the magnetically active column by one order of magnitude, either increasing it or decreasing it depending on whether the magnetic field is parallel or antiparallel to the angular velocity vector of the disc, respectively.  This trend has not yet been reflected  in existing numerical simulations of MRI-induced turbulence \citep[e.g.][]{SS02a, SS02b}. We argue that this is the result of these simulations not yet probing the Hall-dominated region of parameter space, where these effects are expected to be significant (WS11). 

Fig.~\ref{fig:MRI2} compares the growth rate of the most-unstable MRI modes as a function of the magnetic field strength, for different radii in a minimum-mass solar nebula disc and assuming that ions and electrons are the only charge carriers. Note that in all cases the growth rate initially increases with $B$, then levels at $\nu_{\rm max} \sim 0.75$ (the growth rate associated with ideal--MHD perturbations in a Keplerian disc) and, finally, drops abruptly at the field strength for which the wavelength of the perturbations is of the order of the tidal scale-height of the disc. This generally occurs when the ratio of the magnetic pressure to gas pressure at the midplane is of order unity (or $a_{\rm 0} \lesssim 1$;   \citealt{BH91}). However, calculations by \citet{SW03} and WS11 show that when $\eta_{\rm H} < 0$ the instability may grow for super-equipartition (supra-thermal) fields. This is the result of Hall diffusion in this case extending the unstable wavelengths to shorter values. 

These results demonstrate that the MRI grows over a wide range of magnetic field strengths once dust grains have settled out of the gas phase, even in the inner regions of weakly-ionised discs. Note also that the maximum field strength for which perturbations grow is inversely related to the radius. This is expected, as both the midplane gas density $\rho_{\rm 0}$ and the isothermal sound speed $c_{\rm s}$ decrease with the radial location. As a result, the value of $a_{\rm 0} \equiv B_{\rm 0}/[(4 \pi \rho_{\rm 0})^{1/2} c_{\rm s}]$ associated with a particular field strength increases with $r$, causing the instability to be damped at progressively weaker fields as the radius increases.

 \begin{figure*}
\centering
\includegraphics[width=5.9in]{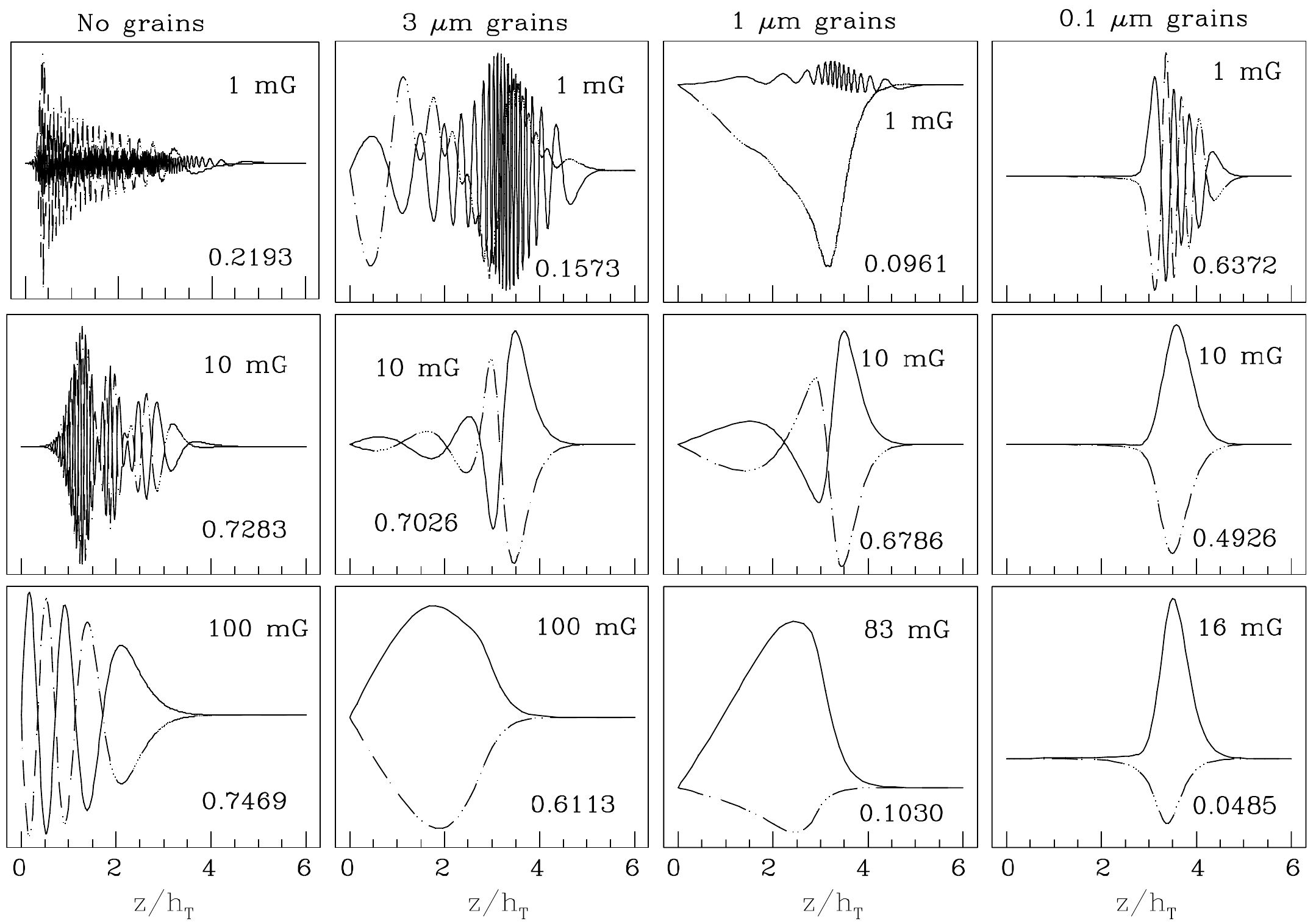}
\caption{As per Fig.~\ref{fig:MRI1} as a function of the presence and size of dust grains suspended in the gas phase, for a minimum--mass solar nebula disc at $r = 5$ AU. All diffusivity components have been incorporated in the ionisation balance, and $B_z > 0$. The leftmost column shows the magnetic field perturbations when the grains have settled out of the gas phase. The remaining ones display (from left to right) the corresponding results when a population of single-sized grains of radius $a_{\rm g} = 3$, $1$ and $0.1 \ \mu$m, respectively, remains suspended with the fluid. The grains constitute a constant fraction (one percent) of the total mass of the gas. 
The lowest row shows the maximum field strength for which the perturbations grow (for $a_{\rm g} = 1$ and $0.1 \ \mu$m), or the results for $B = 100$ mG (for the no grains and the $a_{\rm g} = 3 \ \mu$m cases). Note that when the grains are small ($a_{\rm g} = 0.1 \ \mu$m) the layer below $\sim 2.5$ scale-heights is magnetically inactive. When $B_z > 0$, however, this effect is restricted to a finite range of grain sizes. Once the grains have aggregated to $\sim 1 \ \mu$m or more, the midplane is again unstable to the MRI (see also \citealt{SW08} and WS11). }
\label{fig:MRI3}
\end{figure*}

The magnetic activity of discs is also strongly dependent on the presence and size distribution of dust grains mixed with the gas. This is the result of rapid recombinations taking place on their surfaces, as well as their relatively large cross-sections, which causes them to decouple from the magnetic field at densities for which other  -- smaller -- particles are still well tied to it. The vertical structure and growth of the fastest-growing MRI modes as a function of the presence and size of dust grains suspended in the gas phase is illustrated in Fig.~\ref{fig:MRI3} for the minimum-mass solar nebula model at $r = 5$ AU. The leftmost column of the figure shows the magnetic field perturbations when the grains have settled out of the gas phase, or have aggregated to sizes significantly larger to those shown. The other columns display the corresponding results when a population of single-sized grains, of the radius ($a_{\rm g}$) indicated at the top of each column and constituting a constant (one percent) fraction of the mass of the gas, remains suspended with the fluid. All the magnetic diffusivity components have been incorporated in the calculations and $B_z > 0$. The lowest row shows the maximum field strength for which the perturbations grow (for $a_{\rm g} = 1$ and $0.1 \ \mu$m), or the results for $B = 100$ mG (for the no grains, or $a_{\rm g} = 3 \ \mu$m cases). Note the extended magnetically-inactive zone adjacent to the disc midplane when the grains are small (sub-micron sized). In this case, severe magnetic diffusion in the disc interior prevents the magnetic field from coupling to the gas below $\sim 2.5$ scale-heights. When the grains have aggregated to about a micron in size, however, the disc midplane remains magnetically coupled for all the field strengths for which the instability grows ($B \lesssim 100$ mG, see the column for $a_{\rm g} = 1 \ \mu$m in Fig.~\ref{fig:MRI3}). In contrast, the MRI grows for $B \lesssim 1$ G when no grains are present at this radius\footnote{The MRI grows for $B \lesssim 8$ G (at 1 AU) and $B \lesssim 250$ mG (at 10 AU) when dust grains are absent \citep{SW05}.} \citep{SW05}.  These results hold when the magnetic field is aligned with the rotation axis of the disc, rendering $\eta_{\rm H} > 0$. When the field is counter-aligned to it, however, only a small fraction of the disc is active even after the grains aggregate, unless the field is in the range of 20 -- 80 G (WS11). 
\begin{figure}
\centering
\includegraphics[width=3.0in]{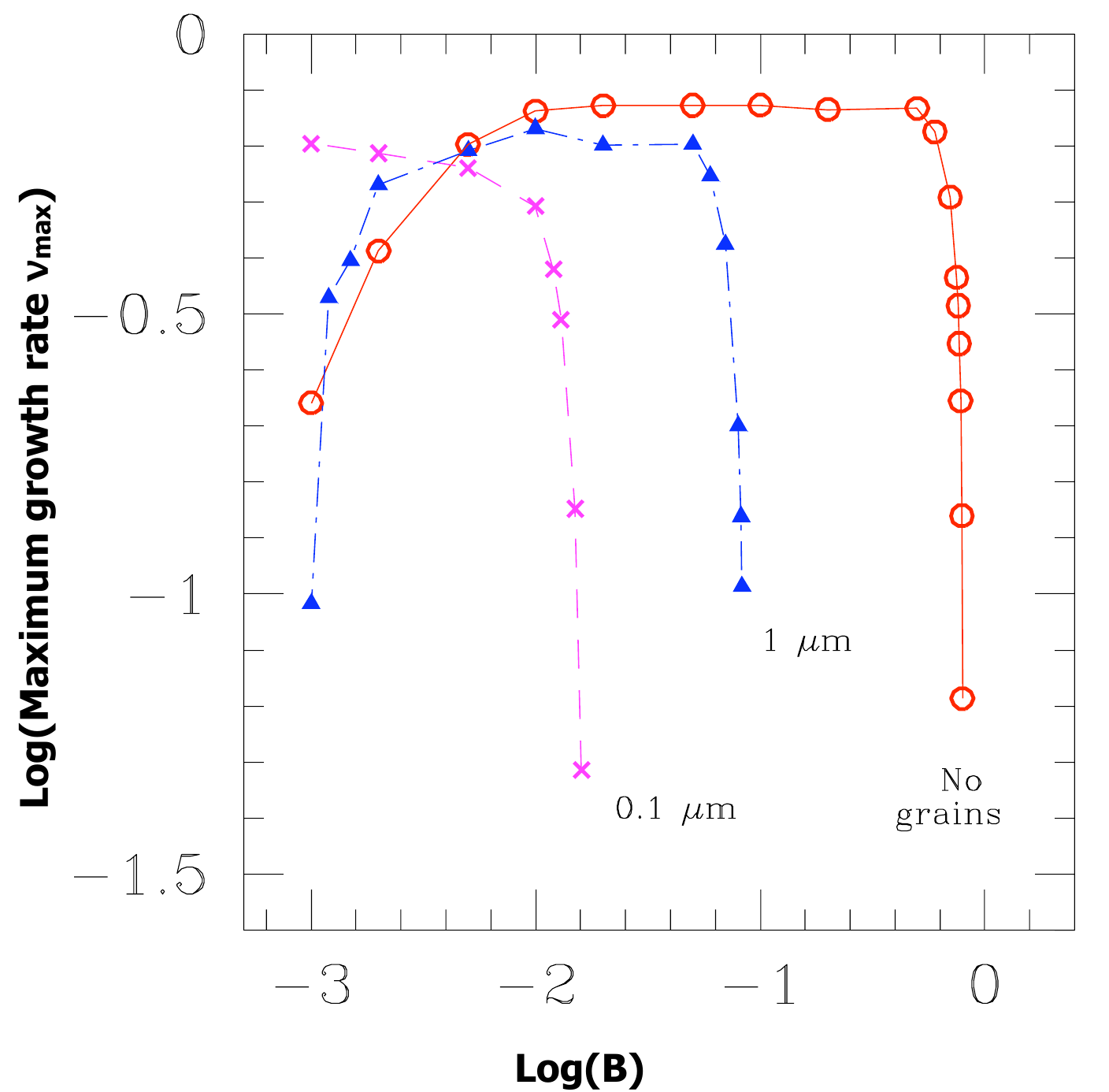}
\caption{As per Fig. \ref{fig:MRI2} for $r = 5$ AU and different assumptions regarding the presence and size of dust particles suspended in the gas. The range of field strengths over which the MRI operates is more restricted as the grains are smaller, given the  associated drop in the field-matter coupling for a particular $B$ as the grain size goes down.}
\label{fig:MRI4}
\end{figure}

Fig.~\ref{fig:MRI4} displays the growth rate of the fastest-growing MRI perturbations at $r = 5$ AU, for different assumptions regarding the presence and size of dust grains mixed with the gas and $B_z > 0$.  The range of field strengths over which the MRI operates is smaller as the grain size diminishes. This is expected, given the reduction in magnetic coupling for a particular $B$ as the dust particles are smaller. Note, however, that the stabilising effect associated with the presence of small grains is not strong enough to completely suppress the perturbations, even when small particles ($0.1 \ \mu$m in size) are present. In this case unstable modes are still able to grow over a finite section of the disc away from the midplane and for a restricted range of field strengths.

All the results discussed so far in this section were obtained assuming that the surface density profile of the disc is that of the minimum-mass solar nebula model, $\Sigma(r) = 1700\  r_{\rm AU}^{-3/2}$ g cm$^{-2}$, where $r_{\rm AU}$ is the radial distance from the central object measured in astronomical units. This yields a surface density of $\sim 150$ g cm$^{-2}$ at 5 AU and $\sim 50$ g cm$^{-2}$ at 10 AU. However, the surface density profile  in real discs may well be different from these inferred values. In fact, recent observations by \citet{Kitamura02a} and \citet{AW07} seem to imply that the surface density declines more gradually with  radius than predicted by this model. Moreover, the actual surface density in the disc inner regions may be smaller than that of a minimum-mass solar nebula disc \citep{AW07}. This would facilitate a deeper penetration of the ionising sources into the disc, and would result in an increased ionisation fraction closer to the midplane. This effect would, in turn, modify the properties of MRI-unstable modes in this region and, more generally, the presence and configuration of the magnetically dead zone in the disc interior. On the other hand, protostellar discs may have surface densities roughly up to $\sim 50$ times the values implied by the minimum-mass solar nebula model without becoming gravitationally unstable. An increased surface density will result in a more extended section of the disc interior being shielded from the ionising effect of X-rays and cosmic rays, but the magnetically active regions closer to the surface are not likely to be substantially modified. 

\section{Combined radial and vertical angular momentum transport}
\label{sec:combined}
The redistribution of angular momentum in real protoplanetary discs is likely to involve \emph{both} the radial and vertical transport mechanisms discussed in the previous two Sections. Although there have been some attempts in the literature to calculate disc models in which these processes coexist \citep[e.g.][]{CF00, OL01}, these studies have not yet linked the generation of radial angular-momentum transport with its origin in magnetic effects via MRI-induced turbulence. In order to devise a method to explore the possible joint operation of these two processes, we make use of the fact that 
the MRI is generally suppressed when the ratio of the Alfv\'en to sound speed $a$ 
is of order unity \citep{BH91}. The instability is expected to develop into full turbulence when $a \ll 1$, provided that the field-matter coupling is sufficient to sustain MHD-driven processes. On the other hand, wind-driving discs require an ordered, relatively strong field (e.g. consistent with $a \lesssim 1$) for the wind to be launched. Since the value of $a \equiv B /[(4 \pi \rho)^{1/2} c_{\rm s}]$ \emph{increases with height} on account of the strong decline in density away from the midplane, this suggests that in principle it is possible for both forms of transport to operate at the same radius, with the MRI transporting angular momentum radially outwards over a finite section of the disc close to the midplane, and vertical transport operating at higher $z$. There may also be a magnetically-inactive (dead) zone across the disc midplane, where the magnetic diffusivity is large enough to prevent the magnetic field from effectively coupling to the fluid and the disc shear. 

The above possibility was first explored by \citet{SKW07a} under the assumption of pure ambipolar diffusivity and constant field-matter coupling. The obtained, preliminary, results seem to indicate that there is little room for the joint operation -- at the same radius, but at different heights -- of these two mechanisms. This is because for this to occur the parameter $a$ must satisfy two conditions: First, it must be small enough near the midplane to allow MRI modes to grow over a finite section of the disc. Second,  it has to be sufficiently large near the surface for the wind to be launched. Physically-viable wind solutions also satisfy $2 \Lambda a^2 > 1$ (WK93, KSW10). In practice, these constraints 
seem to require comparatively strong magnetic fields at the disc midplane (consistent with $a_{\rm 0} \gtrsim 0.5$), 
coupled with low (and decreasing with $z$) values of the field-matter coupling parameter, for both transport mechanisms to operate in the described fashion. These conditions may be difficult to satisfy over extended radial locations in real discs. Despite these initial results, I emphasise that it is crucial to conduct more detailed modelling, incorporating a realistic diffusivity and field-matter coupling profiles, in order to fully explore the relationship between radial and vertical angular-momentum transport in protoplanetary discs.

\section{Closing remarks}
\label{sec:planets}

In this paper I have examined the magnetic activity of protostellar accretion discs, the properties of the field-matter diffusivity in weakly ionised environments, and the two most generally applicable mechanisms for angular momentum transport in these objects: MRI--driven magnetic turbulence and winds launched centrifugally from the disc surfaces. In the first process, the MRI redistributes angular momentum radially outwards via Maxwell stresses associated with small-scale, disordered magnetic fields for which the ratio of the Alfv\'en speed to the sound speed is much smaller than unity ($a \ll 1$, a `weak' field\footnote{Note that, in this scenario, the fluctuating component of the field may typically be stronger than the mean field \citep[e.g.][]{SITS04}.}). The second mechanism transfers angular momentum from the disc material to the wind through the action of a large scale, ordered field (satisfying $a \lesssim 1$ already at the disc midplane; a `strong' field). In this latter scenario, magnetic turbulence is expected to be suppressed as the MRI-unstable modes do not fit within the thickness of the disc (e.g.~WK93).  

The low conductivity of the gas, particularly in the inner regions of the disc, has to be carefully accounted for when modelling these processes. 
The results indicate that, despite the weak ionisation, the field is able to couple to the gas and shear for fluid conditions thought to be satisfied over a wide range of radii in these discs.  When the Hall diffusivity dominates, however, the magnetic activity depends critically on the magnetic field polarity. The likelihood of a joint operation of the two angular-momentum transport mechanisms was briefly discussed. Further work, incorporating realistic fluid conditions, is necessary in order to fully explore this important possibility. 

I close with a summary of the main points discussed in the paper.

\subsection*{\em Magnetic activity}

\begin{enumerate}
\item Angular momentum redistribution (hence, accretion) in protostellar discs is likely to be mediated by the action of magnetic fields through the following two processes:  MHD-induced turbulence (via the magnetorotational instability -- MRI) or magnetocentrifugal jets accelerated from the disc surfaces. 

\item In protostellar discs, the low ionisation fraction makes it imperative to fully account for departures to the ideal-MHD limit when evaluating the ability of the magnetic field to drive these processes. 

\item Three diffusivity regimes can be identified in weakly-ionised environments: Ambipolar, Hall and Ohm. Ambipolar (Ohm) diffusivity dominates in low-density (high-density) regions endowed with a strong (weak) magnetic field, so that the charges are primarily tied to the magnetic field (to the neutrals). The Hall regime is dominant for intermediate densities and field strengths, and is characterised by a varying degree of coupling of different charged species with the magnetic field: Small charges (such as electrons) are coupled to the field, whereas more massive ones (e.g.~ions and dust grains) are tied to the neutral component of the fluid.

\item In a minimum-mass solar nebula disc, the ambipolar, Hall and Ohm regimes are dominant in the disc interior on radial scales $r \gtrsim 10$ AU, $\sim 1 - 10$ AU, and $\sim 0.1 - 1$ AU, respectively.

\item When the Hall diffusivity is important, the magnetic response of the disc is no longer invariant under a global reversal of the magnetic field polarity (e.g.~\citealt{WN99}; SKW11).

\setcounter{mypapers}{\value{enumi}}
\end{enumerate}

\subsection*{\em Magnetocentrifugal Jets}

\begin{enumerate}

\setcounter{enumi}{\value{mypapers}}

\item  Angular momentum can be transported vertically, via the acceleration of magnetocentrifugal jets from the disc surfaces. For this to occur, the magnetic field must be relatively strong, so that the ratio  of the Alfv\'en speed to the sound speed at the midplane is of order unity ($a_{\rm 0} \lesssim 1$), and the magnetic coupling must be sufficiently strong to sustain MHD-driven processes.   

\item Winds accelerated centrifugally from the surfaces of strongly-magnetised discs exhibit the following layered structure, in order of increasing height above the disc midplane.
\begin{enumerate}
\item a {\em quasi-hydrostatic} layer, where the bulk of the matter is located, matter loses angular momentum to the field and strong gradients in the fluid variables take place; 
\item a {\em transition} layer, where the magnetic energy becomes dominant over the fluid thermal energy on account of the drop in density away from the midplane; and 
\item a {\em wind} region, where the field transfers angular momentum back to matter and accelerates it centrifugally. 
\end{enumerate}
In the top-most layer, which constitutes the base of the outflow, all velocity components [$(\mathbf{v} - v_{\rm K}\hat{\mathbf{\phi}})/c_{\rm s}$] are positive and increasing with $z$. In contrast, in the first two (e.g.~within the {\em disc}), the radial velocity is negative, the vertical velocity is small and the azimuthal velocity ($v_\phi$) is sub-Keplerian.

\item We have developed a procedure to explicitly solve for the vertical structure of a radially-localised wind-driving disc model (WK93, KSW10). The solutions are integrated numerically between the midplane and the sonic surface of the flow, and incorporate the strong stratification of the disc  (WK93, SKW11). Furthermore, by matching the solution to the BP82 global self-similar wind models, we are able to obtain solutions that continue to accelerate to the Alfv\'en surface.   

\item In order to facilitate the matching of the localised ({\em wind-driving disc}) solution to the global, self-similar ({\em large-scale wind}) solution, we obtained a library of the global solutions of BP82 for a wide range of the model parameters. The parameter combinations for viable solutions are available in tabular form in the electronic version of SKW11.   

\item We have derived the following parameter constraints, satisfied by physically-viable wind-driving disc solutions (WK93, KSW10).
\begin{enumerate}
\item The flow remains sub-Keplerian below the launching height.
\item A wind-launching criterion is satisfied. This implies the sufficient inclination of the magnetic field lines with respect to the rotational axis of the disc for centrifugal acceleration to take place.
\item The base of the outflow is located above one magnetically-reduced density scale-height.
\item The rate of heating by Joule dissipation at the midplane does not exceed the rate of gravitational potential energy released at that location. 
\end{enumerate}

\item We find that the above constraints lead to four parameter sub-regimes where solutions are found in the Hall limit, and to three in the Ohm limit. Our numerical solutions are in agreement with these constraints. In the regions of parameter space excluded by the constraints, wind-driving disc solutions  cannot be obtained or are unphysical. 

\item For all diffusivity regimes, the ratio of the neutral-ion momentum exchange time to the disc orbital time is larger than unity ($\Upsilon \gtrsim 1$). This is a fundamental property of the wind solutions considered here.

\item Increasing the relative contribution of the Hall diffusivity term, while keeping the remaining model parameters unchanged, results in a higher location of the sonic point, and in a smaller magnetically-reduced density scale-height. Correspondingly, the density at the sonic point, and the mass outflow rate, are both reduced. 

\item When the Hall diffusivity is important, both  the properties of the solutions and the regions of parameter space where these  are found are dependent on the magnetic field polarity. In two Hall sub-regimes (Cases $ii$ and $iv$, KSW10; see also Table \ref{table:constraints}) wind solutions exist only for the positive polarity ($B_z > 0$, meaning that the field is aligned with the angular velocity vector of the disc). In the other two (Cases $i$ and $iii$), no viable solutions are found for $\eta_{\rm H} < 0$ when $\upo < 2 \betiabs^{-1}$ (see Fig.~\ref{fig:5_26}). 

\item When solutions with both field polarities exist, the base of the wind and the sonic surface are both located at a higher $z$ in the solutions with $B_z > 0$ (for similar values of all the other model parameters).  

\setcounter{mypapers}{\value{enumi}}
 
\end{enumerate}

\subsection*{\em Magnetorotational instability}

\begin{enumerate}

\setcounter{enumi}{\value{mypapers}}

\item The magnetorotational instability (MRI) is the most promising angular-momentum transport mechanism in discs that are endowed with a relatively weak magnetic field, so that the ratio $a$ of the Alfv\'en speed to the sound speed is $ \ll 1$, provided that the coupling between the magnetic field and the gas is enough to support MHD-driven processes. 

\item In the weakly-ionised environment typical of the inner regions of protostellar discs, the magnetic diffusivity is strong enough to significantly modify the development and properties of the instability. 

\item A local, linear analysis of the instability (WS11) shows that for the same value of the Pedersen diffusivity $\eta_{\rm P}$, the growth rate increases with the Hall diffusivity term, from zero when $\eta_{\rm H} \Omega_{\rm K}/ v_{\rm A}^2 \leq -2$ to $\sim 0.75 \ \Omega_{\rm K}$, the growth rate associated with ideal-MHD conditions, as $\eta_{\rm H}  \rightarrow \infty$.

\item Hall diffusion modifies the extent of the magnetically-active column by one order of magnitude, increasing it (decreasing it) when the magnetic field is parallel (antiparallel) to the angular velocity vector of the disc (WS11). 

\item MRI-unstable perturbations grow in protostellar discs for a wide range of fluid conditions and magnetic field strengths. When dust grains are absent, the MRI grows at r = 1, 5 and 10 AU for $B \lesssim 8$ G, $B \lesssim 800$ mG and $B \lesssim 250$ mG, respectively \citep{SW05}. 

\item At radial distances within a few AU from the central object, a small (submicron-sized) population of dust grains renders the midplane magnetically inactive. When the magnetic field is aligned with the rotational axis of the disc ($B_z > 0$, resulting in $\eta_{\rm H} > 0$), this effect is restricted to a finite range of grain sizes. Once the grains have aggregated to $\sim 1$ micron or more, the midplane is again unstable to the MRI (\citealt{SW08}; WS11). If the field is counter-aligned, however, only a small section of the disc is unstable, even if the grains aggregate significantly, or are absent (WS11).

\setcounter{mypapers}{\value{enumi}}
\end{enumerate}

\acknowledgments

I am grateful to the anonymous referees for useful comments and suggestions that improved the clarity of the paper.

\end{document}